\documentclass[10pt,journal,compsoc]{IEEEtran}
\usepackage{cite}
\usepackage{amsmath,amssymb,amsfonts}
\usepackage{algorithmic}
\usepackage{graphicx}
\usepackage{textcomp}
\usepackage{bmpsize}
\usepackage{xcolor}
\usepackage{lipsum}
\usepackage[colorlinks=false,urlcolor=black,bookmarks={false}]{hyperref}
\usepackage{acronym}
\usepackage{soul}
\usepackage[final]{microtype}
\usepackage{tabularx}
\usepackage{booktabs}
\usepackage{float}
\usepackage{balance}
\usepackage{multirow}
\usepackage{framed}
\usepackage{pifont}
\usepackage{enumitem}
\usepackage{mathptmx}
\usepackage{xparse}
\usepackage{ragged2e}

\renewcommand{\hl}[1]{}

\usepackage{array}
\newcolumntype{L}[1]{>{\raggedright\let\newline\\\arraybackslash\hspace{0pt}}m{#1}}
\newcolumntype{C}[1]{>{\centering\let\newline\\\arraybackslash\hspace{0pt}}m{#1}}
\newcolumntype{R}[1]{>{\raggedleft\let\newline\\\arraybackslash\hspace{0pt}}m{#1}}
\newcolumntype{P}[1]{>{\raggedright\let\newline\\\arraybackslash}p{#1}}

\newcommand{\subcat}[1]{\hspace{0em}{\textbullet~#1}}

\newsavebox{\fminipagebox}
\NewDocumentEnvironment{fminipage}{m O{\fboxsep}}
 {\par\kern#2\noindent\begin{lrbox}{\fminipagebox}
  \begin{minipage}{#1}
 }
 {\end{minipage}\end{lrbox}%
  \makebox[#1]{%
    \kern\dimexpr-\fboxsep-\fboxrule\relax
    \fbox{\usebox{\fminipagebox}}%
    \kern\dimexpr-\fboxsep-\fboxrule\relax
  }\par\kern#2
 }

\NewDocumentEnvironment{takeaway}{}{\begin{fminipage}{\linewidth}\sffamily\footnotesize\begin{flushleft}}{\end{flushleft}\end{fminipage}}
\acrodef{API}{Application Programming Interface}
\acrodef{SLR}{Systematic Literature Review}

\usepackage{tikz}

\newcommand{\etal}[0]{et~al{.}}

\begin{document}
\title{Don't forget your classics:\\Systematizing 45 years of Ancestry for\\Security API Usability Recommendations}

\author{Nikhil Patnaik, Andrew C Dwyer, Joseph Hallett and Awais Rashid~\IEEEmembership{Member,~IEEE}
  \IEEEcompsocitemizethanks{
    \IEEEcompsocthanksitem N. Patnaik, J. Hallett and A. Rashid are with the Bristol Cyber Security Group, University of Bristol, UK. A.C. Dwyer is with the University of Durham, UK.\protect\\
    {E-mail: \{nikhil.patnaik, joseph.hallett, awais.rashid\}@bristol.ac.uk} \& andrew.dwyer@durham.ac.uk
  }
}


\maketitle
\begin{abstract}
Producing secure software is challenging. The poor usability of security \acp{API} makes this even harder.
Many recommendations have been proposed to support developers by improving the usability of cryptography libraries and \acp{API}; rooted in wider \emph{best practice} guidance in software engineering and API design. In this SLR, we systematize knowledge regarding these recommendations.
We identify and analyze 65 papers spanning 45 years, offering a total of 883 recommendations.
We undertake a thematic analysis to identify 7 core ways to improve usability of \acp{API}.
We find that most of the recommendations focus on helping API developers to \emph{construct} and \emph{structure} their code and make it more usable and easier for programmers to \emph{understand}.
There is less focus, however, on \emph{documentation}, \emph{writing requirements}, \emph{code quality assessment} and the impact of \emph{organizational software development practices}.
By tracing and analyzing paper ancestry, we map how this knowledge becomes validated and translated over time.
We find evidence that less than a quarter of all API usability recommendations are empirically validated, and that recommendations specific to usable security \acp{API} lag even further behind in this regard.


\end{abstract}
\acresetall

\section{Introduction}

Programming is hard to do well, and, even more so, securely. Developers frequently combine functions from \acp{API};
but some are notoriously difficult to use correctly~\cite{robillard2009makes,mclellan1998building},
with cryptography and security libraries often singled out as being particularly obtuse~\cite{kamp2014please,nadi2016jumping}. Strategies ranging from Gamma~\etal{}'s design patterns~\cite{gamma1993design} to the design principles of the 1975 Saltzer \& Schroeder paper on computer security~\cite{saltzer1975protection} have been influential on software engineering practices and, we find, hold a lot of influence over security API design recommendations as well. We investigate how such strategies may have informed recommendations for designing \acp{API}, especially those \acp{API} that provide security and cryptographic functionality.

Over the last 10 years, to help \ac{API} designers produce security \acp{API} that are more usable, various papers have proposed \emph{usability guidelines, principles and recommendations}~\cite{brown2017finding,green2016developers,mendoza2018mobile,mindermann2018rust,patnaik2019usability}\textemdash hereafter \emph{recommendations}.
The field of security \ac{API} usability is, however, relatively new;
with its recommendations building upon work spanning the past 45 years.
Recommendations such as those proposed by Green \& Smith in 2016~\cite{green2016developers}, build upon work on general \ac{API} design by Joshua Bloch from 2006~\cite{bloch2006design}.

Likewise, Bloch~\cite{bloch2001effective} references the design patterns of Gamma~et~al{.}~\cite{gamma1993design}.
This form of ancestry tracing offers a means to systematize the knowledge that inform recommendations for improving usability of security \acp{API}; providing a deeper understanding of current areas of focus, how these have been validated, and where more evidence may be required.
Although previous studies have highlighted existing guidance available to developers~\cite{stylos2007mapping}, no work, to date, has systematized knowledge across 45 years, traced ancestral relationships, as well as the impact of such ancestry on current recommendations for security API usability.



Our \ac{SLR} begins with 13 papers that provide \emph{Security API designer recommendations}. We trace and analyze their ancestry by identifying 883 recommendations in 65 papers across 45 years (Figure~\ref{tab:papers}). These include papers offering general \ac{API} design recommendations (\emph{API designer}), those providing general security best practice (\emph{Security engineering}), and broader software engineering design guidance and recommendations (\emph{Software engineering}).
We categorize recommendations in this corpus and analyze their ancestral chains in order to investigate three research questions:

\textbf{\textit{RQ1: What do current recommendations focus on?}} Using thematic analysis, we developed 36 descriptive themes across 883 recommendations. These 36 themes are consolidated into 7 broad categories.
While many papers have recommended improving documentation as a strategy to assist developers~\cite{bloch2006design,mindermann2018rust,nielsen1994enhancing,beaton2008usability,pane1996usability,zibran2008makes,robillard2009makes,patnaik2019usability,tondel2008security,bloch2001effective}, we are able to excavate how this is reflected in Security API designer usability papers. 
This means we can identify if Security API designer usability recommendations, for example, effectively address \emph{Documentation} or whether they focus more on the \emph{Construction} of \acp{API}.

\begin{table*}
  \caption {Count of Recommendations (Papers) analyzed in the study, broken down by category of paper.}
  \centering
  \footnotesize
  \begin{tabular}{
  >{\RaggedRight}p{\dimexpr 0.16\linewidth-2\tabcolsep}
  p{\dimexpr 0.57\linewidth-2\tabcolsep}
  p{\dimexpr 0.08\linewidth-2\tabcolsep}
  p{\dimexpr 0.19\linewidth-2\tabcolsep}}
    \toprule
    Paper Category & Description & Count & Papers \\
    \midrule
    Security API designer recommendations & Literature that explicitly provides recommendations for improving the usability of security \acp{API} through design. & 84 (13) & \cite{green2016developers, mindermann2018rust, mendoza2018mobile, brown2017finding, patnaik2019usability, votipka2020securitymistakes, oneill2018securesocketAPI, gutmann2002cryptosoftware, oliveira2018APIblindspots, acar2017comparing, egele2013cryptomisuse, georgiev2012mostdangerouscode, meng2018securecodingpracticesjava}\\
  \addlinespace
    API designer recommendations & Literature that focuses on \acp{API}, which may include limited elements of general practice that permits `good' security, but does not explicitly attend to security itself. & 285 (15) & \cite{bloch2006design, henning2007api, clarke2003using, robillard2009makes, mclellan1998building, stylos2007mapping, zibran2008makes, robillard2011APIlearningobstacles, rivieres2004lines, beaton2008usability, nino2009introducing, grill2012methods, bloch2001effective, pane1996usability, ko2004six} \\
  \addlinespace
    Software engineering recommendations & Literature that is around generic software engineering and best practices in the form of recommendations. & 207 (17) & \cite{gamma1993design, myers2016improving, nielsen1994enhancing, cwalina2008framework, green1989cognitive, green1996usability, kannampallil2006handling, fowler2018refactoring, hoffman1990criteria, molich1990improving, holcomb1989amalgamated, polson1990theory, carroll1992getting, nielsen1996usability, heninger2012mining, Ko2004SoftwareErrorsFramework, Smith1982StarUI} \\
  \addlinespace
    Security engineering recommendations & Literature in software engineering and computer security that explicitly make related recommendations but does not specifically focus on API security. & 307 (20) & \cite{OWASPDevGuide, assal2018security, MicrosoftSDL, BSIMM, OWASPSAMM, SeaCord2018SecurePractices, saltzer1975protection, fahl2012eve, ghafari2017security, gutmann1999design, saltzer1974protection, tondel2008security, mead2005security, haley2008security, bostrom2006extending, CLASP2005ApplicationSecurity, meier2006web, lipner2004trustworthy, apvrille2005secure, gorski2018developers} \\
    \midrule
    Total & & 883 (65) & \\
    \bottomrule \\
  \end{tabular}
  \label{tab:papers}
\end{table*}


\textbf{\textit{RQ2: How, and to what extent, have various recommendations been validated?}}
Through a review of recommendations made by different paper types, we find that less than a quarter (across the whole corpus) have been empirically validated. Only 3 papers from Security API designer guidance fall within this category. We also identify which of the 7 broad categories seem to receive greater focus from the research community regarding empirical validation and where potential gaps may lie. 

\textbf{\textit{RQ3: What are the implications of this coverage, in terms of ancestry and their validation, for the emerging set of Security API designer recommendations?}}
In developing ancestries for 13 identified Security API designer papers, we find extended ancestry chains for almost half of these papers, however empirical validation is limited within those chains. We also explore how these ancestries develop---by addressing usability challenges arising from \acp{API} relating to particular languages, specific security problems pertaining to particular applications, or via experiences from developing security analysis and verification tools.

The systematization afforded by our investigation of these three RQs leads to a number of key insights:

\emph{How recommendations propagate over time.} 

`Classics' in the field were identified, such as Saltzer \& Schroeder~\cite{saltzer1975protection}, Nielsen~\cite{nielsen1994enhancing}, and Gamma~\etal{}~\cite{gamma1993design}, who publish recommendations to improve usability through design in software engineering and computer security. Over time, these recommendations were used, reused and adapted to address challenges in niche fields, as well as being empirically validated. Empirical validation is the scientific method of verifying a theory through thorough experimentation. Through empirical validation recommendations are able to gain stronger supporting research evidence.




\emph{The need for deepening the focus on all facets of usability.} Our analysis shows that  API designer and Security API designer papers pay more attention to the technical features of software and \acp{API}. In contrast, Software engineering papers focus strongly on improving the developer's understanding of the code and Security engineering papers address testing of software and \acp{API}. To improve the usability of security \acp{API}, the recommendations the community follow should place an equally strong focus on other themes that also influence their usability.

\emph{The endurance and importance of the Classics~\cite{saltzer1975protection, bloch2001effective, gamma1993design, nielsen1994enhancing}} 
Almost all the Security API designer papers derive their recommendations from well-used `classics'. This is seen in the ten usability recommendations proposed by Green \& Smith that are influenced by the works of both Bloch~\cite{bloch2006design} and Gutmann~\cite{gutmann2002cryptosoftware}, two different ancestry chains originating from Gamma~\etal{}~\cite{gamma1993design} and Saltzer \& Schroeder~\cite{saltzer1975protection}.

The process of tracing the origins of our security API design recommendations introduced us to these works. At this point we asked questions such as:

\begin{itemize}
\item How does a paper written in 1975 by Saltzer \& Schroeder~\cite{saltzer1975protection} stand the test of time and show up in papers written in 2020?
\item Why are these papers still relevant to security API design today?
\item How did these recommendations transition from one community to another? From software engineering to computer security?
\end{itemize}

Evolution appears to occur when faced with a challenge. Saltzer \& Schroeder's design principles were adapted by Gutmann~\cite{gutmann1995cryptlib} to address the challenge of designing secure cryptographic APIs. Bloch needed to adapt his work from addressing the challenges of usage in Java programming to the challenges of API design~\cite{bloch2001effective, bloch2006design}. The reason these recommendations evolved is because they were flexible enough to address arising challenges of security engineering and security API design, while maintaining an actionable element. 

 To provide a stronger empirical footing for the Security API designer research community, there is a need for a concerted effort to empirically validate prior recommendations in their ancestral chains, as well as to refine them and identify where gaps may lie. Works such as Saltzer \& Schroeder~\cite{saltzer1975protection} and Gamma~\etal{}'s design patterns~\cite{gamma1993design} have stood the test of time and we see their implicit influence across the field. Security API designer research needs to engage more explicitly with these classical works to both strengthen its foundations and improve the diversity and breadth of its ancestral links.

Our SLR results in a set of \emph{\textbf{8 meta-recommendations}} which summarize 45 years of design guidance targeted at software engineering and computer security.

\section{Systematic Literature Review}
\label{sec:slr}



\subsection{Identifying seed-papers}

To identify relevant work, we used a \emph{snowball method}~\cite{wohlin2014snowball}.  This required us to first identify \emph{seed papers}, upon which our analysis could be based. By forward and backward snowballing from these seeds, we ensure we found connections between all recommendations emergent from these seeds.

\subsubsection{Identifying Papers}
\label{sec:search-terms}


\newcommand{\fakesection}[1]{\noindent\textbf{\itshape #1.}}

\fakesection{Step 1: Online Search}

We used Google~Scholar and IEEExplorer with the following search terms to select papers that offer Security API designer recommendations: 

\begin{center}
  \ttfamily
  API $\binom{\text{\ttfamily Usability|Guidelines|}}{\text{\ttfamily Principles|Design|Librar(y|ies)}}$? (Security)?
\end{center}
\newpage
\fakesection{Step 2: Review of relevant journals and conferences}

We reviewed six relevant venues from their first issue to their latest available (September 2020) and added any paper that appeared to offer Security API designer recommendations to our initial set. We reviewed the following venues as they represented the leading security and software engineering venues from the ACM, IEEE, and USENIX communities:
\begin{itemize}
\item {IEEE Transactions on Software Engineering (TSE)}, \item {IEEE Symposium on Security \& Privacy (Oakland)}, \item {International Conference on Software Engineering (ICSE)},
\item {USENIX Security Symposium (USENIX)},  \item {International Symposium on Usable Privacy \& Security (SOUPS)}, and \item {ACM Conference on Computer and Communications Security (CCS)}.
\end{itemize}
Papers from other venues were identified through the snowballing process.
We selected 45 papers that provided Security API designer recommendations.

\fakesection{Step 3: Selecting Relevant Work}

From our initial selection, each paper was reviewed independently by 3 authors.
During review, we used the following inclusion and exclusion criteria to decide whether a paper would be included in our seed-list or not.

\noindent\textbf{Inclusion:}

\begin{itemize}
\item The paper gave recommendations about improving an \ac{API} or programming interface.
\item The recommendations aimed to improve the usability of the security API.
\item The \ac{API} was designed to be used by programmers building an application, rather than end-users using a program for security (e.g. to accomplish PGP encryption).
\end{itemize}

\noindent\textbf{Exclusion:}

\begin{itemize}
\item The recommendations were not about \acp{API}, but rather a technology an API might wrap (e.g. the use of various cryptographic modes such as ECB, or the benefits of certain cryptographic algorithms).
\item The recommendations were targeted at improving the engineering quality of an API rather than security directly.  Whilst a secure API is often a well engineered API, the recommendations did not focus on security but rather broader engineering concerns (e.g., reducing an API's size to a few clear methods may reduce confusion, and be easier to verify---but unless the paper explicitly stated that this was for security, it would not be included). 
\item The recommendations must discuss an API---several papers gave security recommendations for configuration management which were similar to recommendations for securing \acp{API}; but these papers focused on advising IT workers who maintain these systems and did not describe \emph{programming} interfaces.
\item The recommendations given were too generic to offer any meaningful advice (e.g. ``an API must be secure''---we agree, but this recommendation offers no advice on how to achieve this).
\end{itemize}
Through the inclusion and exclusion criteria, 13 Security API designer \emph{seed papers} were used in our snowballing process.


\subsubsection{Snowballing}
\fakesection{Step 4: Snowballing}


We performed backward snowballing to trace the ancestry of the recommendations presented by our Security API designer seed papers, and we also performed forward snowballing on every paper along the ancestry chain to see if they were validated by other work~\cite{wohlin2014snowball}. At the end of this step we had 156 papers.

\fakesection{Step 5: Identifying `actionable' recommendations from the snowball process}

We sought papers that provide specific steps to improve \acp{API}, rather than general engineering guidance.
From the 156 papers offering recommendations through the initial search and snowballing, we narrowed our recommendations to those that are \emph{actionable}---
that is they detail specific and clear steps to improving the usability of an API, such as:

\fakesection{Improved Usability}
    \begin{quote}
        ``Easy to use, even without documentation: Developers like end-users do not like to read manuals before getting started. If the API is not self-explanatory or worse gives the false impression that it is, developers will make dangerous mistakes.''~\cite{green2016developers}
    \end{quote}

\fakesection{Offered Guidance}
    \begin{quote}
        ``Economy of mechanism: Keep the design as simple and small as possible. This well-known principle applies to any aspect of a system, but it deserves emphasis for protection mechanisms for this reason: design and implementation errors that result in unwanted access paths will not be noticed during normal use (since normal use usually does not include attempts to exercise improper access paths). As a result, techniques such as line-by-be inspection of software and physical examination of hardware that implements protection mechanisms are necessary. For such techniques to be successful, a small and simple design is essential''~\cite{saltzer1975protection}
    \end{quote}
However, those that were too \emph{general} (i.e. not actionable), such as
describing a general principle that should be taken into account to improve the usability of an API without detailing specifics about how that principle should be implemented were excluded, such as:

\fakesection{Developing General Principles}
\begin{quote}
    ``If it’s hard to find good names, go back to the drawing board.''~\cite{bloch2006design}
\end{quote}

\fakesection{Directions on Design}
\begin{quote}
    ``Offer Meaningful Options. The most crucial aspect of a security warning is to offer meaningful options to get out of the situation that triggered the warning.''~\cite{gorski2018developers}
\end{quote}
From the 156 papers, we identified 65 papers providing 883 actionable recommendations for improving usability and security. 
We also identified 91 papers, that provided more general guidance. 
The actionable papers were taken forward for further analysis, and the 91 general papers were kept to show the connections between papers (i.e. where they had influenced or validated actionable guidance), and to describe the ancestry of API recommendations.

\fakesection{Step 6: Deriving Paper Types}

From our 65 actionable papers, 2 authors allocated each paper to one of four paper types, as shown in Table~\ref{tab:papers}, based on an inductive process derived from the papers themselves. This takes the first paper type of `Security API designer' in addition to three others. This process offers a high-level overview of what each paper category broadly addresses and helps us to understand how recommendations propagate against different communities.

\subsubsection{Analyzing Recommendations}\label{sec:analysis}

\begin{table*}\footnotesize\centering
\caption{Codebook showing \textbf{Categories} and \subcat{Descriptors} for the recommendations identified}
\begin{tabular}{p{\dimexpr 0.25\linewidth-2\tabcolsep}p{\dimexpr 0.745\linewidth-2\tabcolsep}}
\toprule
Code & Description \\
\midrule
\textbf{Assessment} & \textbf{The quality and testing of software and \acp{API}.}\\\addlinespace
\subcat{Quality Engineering} & The development, good practice, and management of software and \acp{API}.\\\addlinespace
\subcat{Quality Assurance} & The methods and tools used to assess and audit software and \acp{API}. \\
\midrule
\textbf{Construction} & \textbf{The technical features of software and \acp{API}.}\\\addlinespace
\subcat{Abstraction} & Expressing different components of software and \acp{API} through abstraction.\\\addlinespace
\subcat{Access Validation} & Complete Mediation---Checking for access.\\\addlinespace
\subcat{Code} & Any code or data involved in the construction of software and \acp{API}.\\\addlinespace
\subcat{Error Handling} & How software and \acp{API} deal with errors.\\\addlinespace
\subcat{Economy of Mechanism} & Ensuring minimalist and simple design of software and \acp{API}.\\\addlinespace
\subcat{Open Design} & Ensuring that it is clear what the design is.\\\addlinespace
\subcat{Technical Specifics} & Any element not covered by the other `Construction' descriptors.\\\addlinespace
\subcat{Durability} & How software and \acp{API} develop over time, are maintained, and can be depreciated. \\
\midrule
\textbf{Default Secure} & \textbf{The different methods and practices to develop security as a fundamental outcome.}\\\addlinespace
\subcat{Bug and Defect Management} & Processes and practices for the handling of bugs and defects.\\\addlinespace
\subcat{Fail-Safe Default} & How does software or an API ensure that it will always provide, by default, the safest option?\\\addlinespace
\subcat{Secure Architecture} & How software and API architecture is designed with security at its core.\\\addlinespace
\subcat{Compartmentalization} & Least Common Mechanism---Ensuring that things are not unnecessarily shared.\\
\midrule
\textbf{Documentation} & \textbf{Documentation methods and practices.}\\\addlinespace
\subcat{Explain} & How well documentation describes the usage of an API or software.\\\addlinespace
\subcat{Inventory / COTS} & The development of an inventory to record the different components of an API and software.\\\addlinespace
\subcat{Telemetry and Reporting} & Ensuring active collection and recording of data and information.\\\addlinespace
\subcat{Publish and Communicate} & The use of documentation to distribute or to offer information.\\\addlinespace
\subcat{Standardized} & Ensuring that documentation provides cohesive standards.\\\addlinespace
\subcat{Exemplars} & The use of examples (frequently code) to help explain different aspects of software and \acp{API}.\\\addlinespace
\subcat{Guidance} & The development of guidance for users or designers.\\
\midrule
\textbf{Organizational Factors} & \textbf{How organizations respond to developing software and \acp{API} and interface with external factors.}\\\addlinespace
\subcat{Incident Response} & The development of practices to respond to emergencies or incidents from software and \acp{API}.\\\addlinespace
\subcat{Security Practice} & How an organization develops knowledge and practice of security.\\\addlinespace
\subcat{Training} & The delivering of training for organizations and their members.\\\addlinespace
\subcat{Third Party} & How organizations interface with third parties and third party components.\\\addlinespace
\subcat{Regulatory} & Any regulatory, legal, or compliance that an organization does.\\\addlinespace
\subcat{Risk Assessment and Metrics} & Assessing risk and developing metrics to measure it.\\
\midrule
\textbf{Requirements} & \textbf{The development of requirements for software and \acp{API}.}\\\addlinespace
\subcat{Implement Requirements} & The implementation and application of requirements.\\\addlinespace
\subcat{Write Requirements} & The construction, identification, and development of requirements.\\
\midrule
\textbf{Understanding} &	\textbf{How software and \acp{API} come to be understood and practiced by humans.}\\\addlinespace
\subcat{Assist} & Psychological Acceptability---how an API user or API developer deals with the load of programming and techniques to assist developers.\\\addlinespace
\subcat{Drawing Attention} & Highlighting or pointing towards information required for proper or secure use of software and \acp{API}.\\\addlinespace
\subcat{Misuse} & The prevention of an API user or API developer misusing software and \acp{API}.\\\addlinespace
\subcat{Relevant Information} & The provision of information that concerns a particular task or object of study.\\\addlinespace
\subcat{Meaningful Options} & The provision of options that make sense to API users.\\\addlinespace
\subcat{Sufficient Information} & The provision of enough information in order to effectively communicate and provide understanding.\\\addlinespace
\subcat{Validation of Activity} & Providing API users and API developers tools that check their activities.\\
\bottomrule
\end{tabular}
\label{tab:category-code-book}
\end{table*}

\fakesection{Step 7: Categorizing Recommendations}

To understand the different areas of recommendation focus,
we categorized each of our 65 actionable guidance papers.
To alleviate bias from predefined categorization
the analysis followed an inductive, bottom-up approach~\cite{rivas2012coding} to draw out recommendation themes.

\begin{enumerate}
    \item Two authors, in joint discussion, selected 50 recommendations to identify different \emph{codes} in order to build a mutual understanding of the recommendations. An initial \emph{codebook}~\cite{ando2014achieving} with 28 codes was created.
    
    \item Over three iterations, the 883 recommendations were categorized using the initial codebook.
    Additional codes were created to capture the diversity of recommendations identified during the process.
    The initial codebook was updated to include 19 categories and 75 descriptive sub-categories.
    
    \item Through a visual whiteboard mapping discussion between three authors, the number of codes were reduced and amalgamated.
    This resulted in a consolidated codebook with 7 categories, and 36 descriptive sub-categories, as shown in Table~\ref{tab:category-code-book}.
    The mappings of categories onto recommendations was updated using the new codebook.
\end{enumerate}
Most recommendations were assigned a single top-level category and one of its sub-categories.
A sixth ($\frac{162}{883}$, 18\%) were more complex---exhibiting multiple elements for API design, security or general software engineering guidance into one---and so two categories were used. No recommendation required more than two top-level categories. For example,
Pane and Myers recommend supporting novice programmers:
\begin{quote}
    ``Use Signalling to Highlight Important Information''~\cite{pane1996usability}
\end{quote}
This was assigned to the \emph{Understanding: Drawing Attention} category and descriptor sub-category as it is concerned with helping the developer identify relevant information. Later, Pane and Myers also recommend:
\begin{quote}
    ``Help detect, diagnose, and recover from errors''~\cite{pane1996usability}
\end{quote}
This was assigned two categories: \emph{Understanding: Assist} as the recommendation concerns developer assistance to diagnose problems, and \emph{Construction: Error Handling} as it deals with recovery from errors.


\fakesection{Step 8: Validating categorization}

To validate our categories, a random 10\% sample of the recommendations were assessed by an independent coder.
Using \emph{Cohen's $\kappa$} (a common measure of interrater reliability~\cite{cohen1960coefficient}), and using only a single category per item (as Cohen's $\kappa$ only deals with single categorizations per subject), we calculated a $\kappa$ of 0.74 when mapping the categories, and 0.76 when mapping the descriptors---indicating substantial agreement~\cite{landis1977kappa} between coders.

\subsubsection{Validation Analysis}\label{sec:analysis}
\fakesection{Step 9: Assessing how papers relate to each other}



Using the relationships between papers captured in group discussion between three authors, 5 different \emph{ancestor--descendant relationships}---where papers interact with each other's work---were identified. Within the ancestor--descendant relationships, we make a distinction between empirical validation and the 4 remaining relationships. Through empirical validation it is possible to test recommendation effectiveness through experimentation or systematic observation. However, this bar is lower for the other ancestor--descendant relationships. These were assigned by one author and reviewed by another. These are presented in Table~\ref{fig:validation-code-book}. Below are examples of how each ancestor--descendant relationship and empirical validation relate to the literature.

\begin{table}
\caption{Codebook used for describing 5 different kinds of ancestor--descendant relationships between papers, including empirical validation.}
\footnotesize\centering
\begin{tabular}{p{\dimexpr 0.25\linewidth-2\tabcolsep}p{\dimexpr 0.75\linewidth-2\tabcolsep}}
\toprule
Code & Description \\
\midrule
Distillation & Descendant condenses a Ancestor's recommendation to further build upon it by addressing a more specific challenge. \\
\addlinespace
Borrowed & Descendant addresses Ancestor's guidelines at a superficial level either to review or to run an experiment and analyze their results.\\
\addlinespace
Adaptation & Descendant has translated recommendations from a Ancestor, either through re-wording or forming their own recommendations directly derived from the Ancestor. \\
\addlinespace
Comparison & Descendant contrasts Ancestor recommendation with another set of recommendations, perhaps written by another Descendant. Or Descendant compares their recommendation to that of their Ancestor.\\
\addlinespace
Empirical & Descendant has experimented and evaluated through research a Ancestor recommendation. Descendant assesses whether the Ancestor's recommendations is valid.\\
\bottomrule
\end{tabular}
\label{fig:validation-code-book}
\end{table}

\textbf{Distillation}
    An evolution occurs from Bloch's 2001 book \emph{Effective Java}~\cite{bloch2001effective} to his 2006 paper \emph{How to Design a Good API and Why it Matters}~\cite{bloch2006design}.
    The book on Java programming is condensed by the short paper for API usability recommendations---with the latter paper used frequently by Security API designer papers.
    
\textbf{Borrowed}
    Myers and Stylos~\cite{myers2016improving} refer to the ancestry between Grill~\etal{}~\cite{grill2012methods} and Nielsen and Molich~\cite{nielsen1990heuristic}:

    \begin{quote}
    ``Grill~\etal{} described a method where they had experts use Nielsen’s Heuristic Evaluation to identify problems with an API and observed developers learning to use the same API in the lab. An interesting finding was these two methods revealed mostly independent sets of problems with that API.''~\cite{myers2016improving}
    \end{quote}

    \hl{How is this borrowing though? Write a sentence with the word X borrowed Ys recommendations.}
    
\textbf{Adaptation}
    In 2017 Acar~\etal~\cite{acar2017comparing} ran an experiment to evaluate and compare the usability of 5 Python-based cryptographic libraries.
    To evaluate the usability of these cryptographic libraries, Acar~\etal{} molded recommendations from Bloch~\cite{bloch2006design}, and from Green \& Smith~\cite{green2016developers}. For example, Acar~\etal{} state:
    \begin{quote}
    ``We adapt guidelines from these various sources [Bloch~\cite{bloch2006design}, Green \& Smith~\cite{green2016developers}] to evaluate the \acp{API} we examine.''~\cite{acar2017comparing}
    \end{quote}
    Acar \emph{adapts} Bloch and Green \& Smith's recommendations.

\textbf{Comparison}
    Smith~\cite{smith2012contemporary} reflects on the work of Saltzer~\cite{saltzer1974protection} along with the 1975 paper written with Schroeder~\cite{saltzer1975protection}. Here, Smith compares the `principles' of Saltzer \& Schroeder to those of the-then contemporary recommendations in software security.
    \begin{quote}
     ``The following are new---or newly stated---principles compared to those described in 1975.''~\cite{smith2012contemporary}
    \end{quote}
    Smith makes a \emph{comparison} to the principles of Saltzer \& Schroeder.




\textbf{Empirical}
In 2019 Patnaik~\etal{}~\cite{patnaik2019usability} \emph{empirically} evaluate the 10 principles designed to improve usability and security of \acp{API} by Green and Smith~\cite{green2016developers}.
\begin{quote}
 ``An empirical validation of Green and Smith’s principles showing when a principle is not being applied but also identifying issues that Green and Smith’s principles currently do not capture.''~\cite{patnaik2019usability}
\end{quote}

\section{SLR Findings}

Table~\ref{tab:category-code-book} outlines the 7 recommendation categories and 36 descriptor sub-categories.
The 7 categories describe overarching themes and topics about which papers make recommendations.
The descriptor sub-categories offer greater detail within each of the categories.


For example: the \emph{Construction} category captures recommendations about how to structure and build software.
Its \emph{Code} descriptor sub-category identifies a focus on particular programming details.
Bloch, for instance, advises developers to:
\begin{quote}
    ``Return zero-length arrays, not nulls.''~\cite{bloch2001effective}
\end{quote}
In contrast, the \emph{Economy of Mechanism} descriptor identifies the simple code construction to avoid errors---%
found in Nino~et~al{.}'s \emph{``be minimal''}~\cite{nino2009introducing}, Grill~et~al{.}'s \emph{``Do not provide multiple ways to achieve one thing''}~\cite{grill2012methods}, OWASP's \emph{``Keep It Simple, Stupid Principle''}~\cite{OWASPDevGuide} or Saltzer \& Schroeder's \emph{Economy of Mechanism} principle~\cite{saltzer1975protection}; from which we name the descriptor.
We also observe recommendations on how to document code (the \emph{Documentation} category)---typically focused on clear explanation, communication, standardization and exemplars.
Recommendations also assist \ac{API} users' \emph{Understanding} by aligning concepts with their mental models and helping them with the cognitive load of programming (the \emph{Assist} descriptor): for example Ko~et~al{.} recommend:
\begin{quote}
    ``Help programmers recover from interruptions or delays by reminding them of their previous actions''~\cite{Ko2004SoftwareErrorsFramework}
\end{quote}
Other topics in the \emph{Understanding} category include \emph{Drawing Attention} to \emph{Relevant} and \emph{Sufficient} information, as well as providing \emph{Meaningful Options}.


Figure~\ref{tab:categories-counts} shows the number of recommendations in each paper type mapped to each category and descriptive sub-category.
We find that, over the 883 recommendations, the majority concern the construction and structure of code (the \emph{Construction} category, 32\%), as well as helping to make the code easier to comprehend and clear to the developer (the \emph{Understanding} category, 23\%). The remaining recommendations are more or less evenly spread across the 5 remaining categories (around 7--14\%).

\subsection{RQ1: What do current recommendations focus on?}
\begin{takeaway}
\noindent\textbf{Take-Away 1:}
\begin{itemize}
\item API designer and Software engineering papers focus on how to \emph{structure} code and how to make it \emph{understandable}.
\item Practically only Security engineering papers make recommendations about \emph{Organizational Factors}.
\item Security API designer papers do not engage sufficiently with various aspects of \emph{Documentation} or \emph{Understanding: Validation of Activity}, which should be addressed in future work.
\end{itemize}
\end{takeaway}

\begin{figure*}
  \centering\footnotesize
  \newcommand{\tableheader}[1]{\rotatebox{90}{#1}}
  \begin{tabular}{lccccc}
\toprule
Recommendation Category & \tableheader{API designer guidance} & \tableheader{Security API designer guidance} & \tableheader{Software engineering guidance} & \tableheader{Security engineering guidance} & Overall \\
\midrule
Total & 285 & 84 & 207 & 307 & 883 \\
\midrule
Assessment & 20 (7\%) & 8 (10\%) & 18 (9\%) & 76 (25\%) & 122 (14\%)\\
\addlinespace
\subcat{Quality Engineering} & 3 (1\%) & 3 (4\%) & 4 (2\%) & 36 (12\%) & 46 (5\%)\\
\subcat{Quality Assurance} & 17 (6\%) & 5 (6\%) & 14 (7\%) & 40 (13\%) & 76 (9\%)\\
\midrule
Construction & 163 (57\%) & 30 (36\%) & 52 (25\%) & 35 (11\%) & 280 (32\%)\\
\addlinespace
\subcat{Abstraction} & 2 (1\%) & 1 (1\%) & 3 (2\%) & 0 & 6 (1\%)\\
\subcat{Access Validation} & 2 (1\%) & 5 (6\%) & 1 (1\%) & 10 (3\%) & 18 (2\%)\\
\subcat{Code} & 79 (28\%) & 14 (17\%) & 11 (5\%) & 8 (3\%) & 112 (13\%)\\
\subcat{Error Handling} & 11 (4\%) & 1 (1\%) & 1 (1\%) & 0 & 13 (2\%)\\
\subcat{Economy of Mechanism} & 18 (6\%) & 5 (6\%) & 32 (16\%) & 5 (2\%) & 60 (7\%)\\
\subcat{Open Design} & 1 (\textless 1\%) & 1 (1\%) & 0 & 4 (1\%) & 6 (1\%)\\
\subcat{Durability} & 18 (6\%) & 2 (2\%) & 3 (2\%) & 5 (2\%) & 28 (3\%)\\
\subcat{Other} & 32 (11\%) & 1 (1\%) & 1 (1\%) & 3 (1\%) & 37 (4\%)\\
\midrule
Default Secure & 5 (2\%) & 9 (11\%) & 8 (4\%) & 42 (14\%) & 64 (7\%)\\
\addlinespace
\subcat{Bug and Defect Management} & 0 & 1 (1\%) & 2 (1\%) & 6 (2\%) & 9 (1\%)\\
\subcat{Fail-Safe Default} & 0 & 4 (5\%) & 1 (1\%) & 4 (1\%) & 9 (1\%)\\
\subcat{Secure Architecture} & 3 (1\%) & 4 (5\%) & 5 (2\%) & 28 (9\%) & 40 (5\%)\\
\subcat{Compartmentalization} & 2 (1\%) & 0 & 0 & 4 (1\%) & 6 (1\%)\\
\midrule
Documentation & 28 (10\%) & 14 (17\%) & 11 (5\%) & 40 (13\%) & 93 (11\%)\\
\addlinespace
\subcat{Explain} & 10 (4\%) & 3 (4\%) & 8 (4\%) & 2 (1\%) & 23 (3\%)\\
\subcat{Inventory / COTS} & 2 (1\%) & 0 & 0 & 4 (1\%) & 6 (1\%)\\
\subcat{Telemetry and Reporting} & 0 & 0 & 0 & 13 (4\%) & 13 (2\%)\\
\subcat{Publish and Communicate} & 1 (\textless 1\%) & 1 (1\%) & 0 & 9 (3\%) & 11 (1\%)\\
\subcat{Standardized} & 3 (1\%) & 1 (1\%) & 0 & 5 (2\%) & 9 (1\%)\\
\subcat{Exemplars} & 6 (2\%) & 3 (4\%) & 0 & 0 & 9 (1\%)\\
\subcat{Guidance} & 6 (2\%) & 6 (7\%) & 3 (2\%) & 7 (2\%) & 22 (3\%)\\
\midrule
Organizational & 0 & 3 (4\%) & 0 & 51 (17\%) & 54 (6\%)\\
\addlinespace
\subcat{Incident Response} & 0 & 0 & 0 & 5 (2\%) & 5 (1\%)\\
\subcat{Security Practice} & 0 & 1 (1\%) & 0 & 12 (4\%) & 13 (2\%)\\
\subcat{Training} & 0 & 2 (2\%) & 0 & 13 (4\%) & 15 (2\%)\\
\subcat{Third Party} & 0 & 0 & 0 & 1 (\textless 1\%) & 1 (\textless 1\%)\\
\subcat{Regulatory} & 0 & 0 & 0 & 7 (2\%) & 7 (1\%)\\
\subcat{Risk Assessment and Metrics} & 0 & 0 & 0 & 13 (4\%) & 13 (2\%)\\
\midrule
Requirements  & 10 (4\%) & 0 & 6 (3\%) & 55 (18\%) & 71 (8\%)\\
\addlinespace
\subcat{Implement Requirements} & 4 (1\%) & 0 & 0 & 3 (1\%) & 7 (1\%)\\
\subcat{Write Requirements} & 6 (2\%) & 0 & 6 (3\%) & 52 (17\%) & 64 (7\%)\\
\midrule
Understanding & 59 (21\%) & 20 (24\%) & 112 (54\%) & 8 (3\%) & 199 (23\%)\\
\addlinespace
\subcat{Assist} & 36 (13\%) & 6 (7\%) & 52 (25\%) & 2 (1\%) & 96 (11\%)\\
\subcat{Drawing Attention} & 1 (\textless 1\%) & 2 (2\%) & 8 (4\%) & 0 & 11 (1\%)\\
\subcat{Misuse} & 5 (2\%) & 2 (2\%) & 5 (2\%) & 0 & 12 (1\%)\\
\subcat{Relevant Information} & 2 (1\%) & 1 (1\%) & 11 (5\%) & 1 (\textless 1\%) & 15 (2\%)\\
\subcat{Meaningful Options} & 11 (4\%) & 4 (5\%) & 27 (13\%) & 1 (\textless 1\%) & 43 (5\%)\\
\subcat{Sufficient Information} & 1 (\textless 1\%) & 0 & 3 (2\%) & 1 (\textless 1\%) & 5 (1\%)\\
\subcat{Validation of Activity} & 3 (1\%) & 5 (6\%) & 6 (3\%) & 3 (1\%) & 17 (2\%)\\
\bottomrule \\
\end{tabular}

  \caption{Recommendations mapped to category and paper type.  Some recommendations were assigned multiple categories.  All of the recommendations were assigned to at least 1 category.}
  \label{tab:categories-counts}
\end{figure*}

Recommendations, as derived from our literature search and ancestral tracing, tend to favor technical aspects. However, when we break these down by paper type, we see significant variations, enabling us to assess what may currently be missed by various paper types and how Security API designer literature compares to other communities.
Both \emph{API designer} and \emph{Security API designer} paper types offer more recommendations on API \emph{Construction} and its \emph{Code}.
The \emph{Construction} category is associated with 57\% of API designer and 36\% of Security API designer paper types, with \emph{Understanding} covering 21\% and 24\% of the paper types respectively.
As API-related recommendations are likely to deal with the interface with code, it is reasonable to expect these paper types to focus more on \emph{Construction}.
Recent work on recommendations for security API usability~\cite{green2016developers,patnaik2019usability} suggest that we may be witnessing a move to recommendations around improving code usability (\emph{Understanding}), but this is limited by the number of papers in this type (13---see Figure~\ref{tab:papers}).

For recommendations in \emph{Software engineering guidance}, this relationship is reversed. A greater emphasis is placed on \emph{Understanding} (54\%), with a reduction in a focus on \emph{Construction} (25\%).
Security engineering papers also covered different topics.
Unlike other paper types, \emph{Security engineering guidance} focused less on \emph{Construction} and \emph{Understanding} (11\% and 3\% respectively), and instead offered greater attention to other categories such as \emph{Assessment} (25\%) and \emph{Requirements} (18\%).

Recommendations categorized under \emph{Organizational Factors} are almost exclusively derived from Security engineering papers. These recommendations concern how an organization and its developers follow best practice, legal requirements, and incident handling processes for software defects. The relationship between corporate literature and academic literature is present in Security engineering papers, in Tondel~\cite{tondel2008security} and Assal~\cite{assal2018security} (Figure~\ref{fig:SLRAncestryMap}-Appendix). Many of the recommendations in this category come from corporate grey literature (such as Microsoft's SDL~\cite{MicrosoftSDL}, The BSIMM framework~\cite{BSIMM} and OWASP~\cite{OWASPSAMM}).

For example, Microsoft's SDL encourages developers to \emph{``Establish a standard incident response process''}~\cite{MicrosoftSDL} so that there are mechanisms for dealing with software defects when they are inevitably discovered (which we capture under the \emph{Incident Response} descriptor).
BSIMM recommends that organizations should \emph{``educate executives''}~\cite{BSIMM} so that `decision-makers' in an organization are sufficiently knowledgeable about security (\emph{Organizational Factors: Training}).
Interestingly, recommendations focused on organizations were almost exclusively limited to Security engineering papers.
This suggests a greater emphasis in the security community on considering the wider implications for developers and software in organizations and the impact of external contexts on being secure. The recommendations reflect how organizational factors may affect the implementation and design of security \acp{API}.


For Security API designer papers, we find that in the \emph{Construction} category there is a greater emphasis on the \emph{Code} sub-category. This shows a dominance in research on what challenges developers face, when working with Security \acp{API}, on the writing of code and implementing the functions of the API. This challenge is studied in more depth by Georgiev~\etal{} who provide recommendations to mitigate and resolve the issue for various parties~\cite{georgiev2012mostdangerouscode}.
At 17\%, this is the largest percentage of descriptors in this paper type.
There are some descriptor categories where there are no instances, however:
\begin{itemize}
    \item \emph{Documentation}
    \begin{itemize}
    \item Inventory/COTS;
    \item Telemetry and Reporting,
    \end{itemize}
    \item \emph{Organizational Factors}
    \begin{itemize}
    \item Incident Response;
    \item Third Party;
    \item Regulatory;
    \item Risk Assessment and Metrics,
    \end{itemize}
    \item \emph{Requirements}
    \begin{itemize}
    \item Implement Requirements;
    \item Write Requirements,
    \end{itemize}
    \item \emph{Understanding} 
    \begin{itemize}
    \item Sufficient Information.
    \end{itemize}
\end{itemize}
Whereas some of these may be less pertinent to Security API designer papers, our analysis highlights that further consideration of these may be required to increase the breadth and depth of the community's engagement with different aspects pertinent to improving Security API designer usability.



\subsection{\mbox{RQ2: Are we validating recommendations?}}

\begin{takeaway}
\noindent\textbf{Take-Away 2:}
\begin{itemize}
    \item Today's Security API designer recommendations build upon those presented in historical papers. However, across this ancestry, only 22\% of the papers are empirically validated, meaning further work should be conducted to strengthen their foundations.

    \item Of the Security engineering papers that receive empirical validation or are part of a ancestor--descendant relationship, more than half are through an adaptation of recommendations.

    \item Only 3 of the 13 Security API designer paper have been empirically validated.

\end{itemize}
\end{takeaway}

Our analysis identifies the need for a stronger focus on validating the various recommendations across different paper types, but particularly so for Security API designer papers.
To assess the relationships between different recommendations over time, we constructed their ancestry by separating each recommendation's inheritance into five distinct ancestor--descendant relationships (see Figure~\ref{fig:validation-code-book}):
\emph{Distillation}, \emph{Borrowed}, \emph{Adaptation}, \emph{Comparison}, and \emph{Empirical}.



\subsubsection{Validation by Paper Type}

\begin{figure}
  \centering\footnotesize
  \newcommand{\tableheader}[1]{\rotatebox{90}{#1}}
  \begin{tabular}{lcccccc}
\toprule
Validation & \tableheader{API designer guidance} & \tableheader{Security API designer guidance} & \tableheader{Software engineering guidance} & \tableheader{Security engineering guidance} & Overall \\
\midrule
Adaptation & 9 & 2 & 7 & 16 & 34 (52\%) \\
Empirical & 1 & 3 & 8 & 2 & 14 (22\%) \\
Distillation & 7 & 8 & 8 & 1 & 24 (37\%) \\
Comparison & 1 & 7 & 3 & 4 & 15 (23\%) \\
Borrowed & 7 & 2 & 2 & 4 & 15 (23\%) \\
\addlinespace
\multirow{2}{*}{Overall} & 25 & 22 & 28 & 27 & 26 \\
& (39\%) & (33\%) & (43\%) & (42\%) & (40\%) \\
\bottomrule\\
\end{tabular}

  \caption{Rates of different kinds of paper validation for different categories of guideline papers in the literature. The {overall} columns and rows account for single papers being validated multiple times.}
  \label{tab:validation-counts}
\end{figure}



Figure~\ref{tab:validation-counts} summarizes the number of papers associated with empirical validation and the other ancestor--descendant relationships. Full charts of the various relationships we identified through mappings are presented in the Appendix in Figure~\ref{fig:SLRAncestryMap}; these show the full ancestry of the recommendations.

Overall 22\% of the papers engage in empirical validation of prior work. 8 (53\%) Software engineering and 2 (47\%) Security engineering papers are empirically validated, whereas only 3 Security API designer and 1 API designer papers are empirically validated.
Recommendations written more recently, as part of the software engineering and the computer security community, have developed upon some form of ancestor--descendant relationship or empirical validation of Software engineering and Security engineering papers like Nielsen's usability heuristics~\cite{nielsen1994enhancing} and Saltzer \& Schroeder's principles~\cite{saltzer1975protection}.
Though efforts have been made to empirically validate older papers~\cite{nielsen1994enhancing, saltzer1975protection}, contemporary API recommendations are inherited from a large corpus of papers,
where only 22\% are empirically validated.
This raises the need to further understand and validate how API recommendations are built.
Out of 13 Security API designer papers, only 3 papers~\cite{green2016developers, georgiev2012mostdangerouscode, egele2013cryptomisuse} are empirically validated.
As we create further recommendations, we must consider the role of ancestry, and the validation of what it recommends, in order to strengthen the foundations of Security API designer recommendations. Otherwise we have no way of establishing if particular recommendations---and efforts invested in following them---have a material impact on improving the usability of security \acp{API}. We also risk propagating ineffective recommendations over time.

\subsubsection{Which aspects are we validating?}

If certain academic literature is not conducting extensive and in-depth validation of all areas, then which aspects are we validating?
Figure~\ref{tab:validated-category-counts} counts the different recommendation categories broken down by their ancestor--descendant relationship, including empirical validation.
We empirically validate more on \emph{Construction} (45\%) followed by \emph{Understanding} (27\%).
For other ancestral relationships, categories exhibit different rates, but overall these are at the levels we would expect given their relative frequency across different paper types (Figure~\ref{tab:validated-category-counts}).
The software engineering and security communities should focus on forming more empirical and comparison based relationships, as opposed to borrowing and distillation, to best ensure the effectiveness of the recommendations with thorough, repeatable experimentation (see Figure~\ref{tab:validated-category-counts}).



\begin{figure*}
    \centering\footnotesize
    \newcommand{\tableheader}[1]{\rotatebox{90}{#1}}
    \begin{tabular}{lcccccccc}
\toprule
Recommendation Category & \tableheader{Adaptation} & \tableheader{Empirical} & \tableheader{Distillation} & \tableheader{Comparison} & \tableheader{Borrowed} & Overall \\
\midrule
Total & 727 & 166 & 283 & 91 & 148 & 1415 \\
\addlinespace
Construction & 288 (40\%) & 75 (45\%) & 162 (57\%) & 23 (25\%) & 49 (33\%) & 597 (42\%)\\
Documentation & 80 (11\%) & 18 (11\%) & 25 (9\%) & 9 (10\%) & 35 (24\%) & 167 (12\%)\\
Requirements & 61 (8\%) & 3 (2\%) & 11 (4\%) & 1 (1\%) & 8 (5\%) & 84 (6\%)\\
Understanding & 106 (15\%) & 45 (27\%) & 53 (19\%) & 28 (31\%) & 41 (28\%) & 273 (19\%)\\
Assessment & 102 (61\%) & 21 (13\%) & 17 (26\%) & 24 (16\%) & 8 (1\%) & 172 (12\%)\\
Default Secure & 35 (5\%) & 4 (2\%) & 15 (5\%) & 5 (6\%) & 7 (5\%) & 66 (5\%)\\
Organisational and Regulatory Factors & 55 (8\%) & 0 & 0 & 1 (1\%) & 0 & 56 (4\%)\\
\bottomrule \\
\end{tabular}

    \caption{Counts of recommendations that have been empirically validated or part of other ancestor--descendant relationships across the 7 broad category types.}
    \label{tab:validated-category-counts}
\end{figure*}

\subsection{RQ3: Where do Security API Designer Recommendations Come From?}


\begin{takeaway}
\noindent\textbf{Take-Away 3:}
\begin{itemize}
    \item Almost half of the Security API designer papers have a well defined and long ancestry, dating back to 1974.
    \item A distinction between the capacity to validate \emph{abstract} and \emph{concrete} recommendations (derived from experiences with particular tools and applications) exists.
    \item Recommendations derive mainly from `standalone' ancestries, or are processed through Gutmann~\cite{gutmann2002cryptosoftware} or subsequently through Green \& Smith~\cite{green2016developers}.
\end{itemize}
\end{takeaway}


In the development of Security API designer recommendations, there are two broad forms---abstract and concrete---in how they are developed that we identified in the ancestries we analyzed.

First, \emph{abstract} recommendations such as by Green \& Smith~\cite{green2016developers} apply to a number of tools, applications, and contexts.
Second, there are \emph{concrete} recommendations as identified by the ancestries of tools and applications~\cite{egele2013cryptomisuse, mendoza2018mobile}.
These tend to be more tightly focused to a particular tool or application---and therefore offer advice, that as one would expect, focuses more exclusively on \emph{Construction} and \emph{Requirements}.

\fakesection{\emph{Abstract} Recommendation}
    \begin{quote}
        ``Defaults should be safe and never ambiguous.''~\cite{green2016developers}
    \end{quote}

\fakesection{\emph{Concrete} Recommendation}
    \begin{quote}
        ``Client-side validation must be thoroughly tested for consistency with server-side validation logic. WARDroid can help in identifying potential inconsistencies''~\cite{mendoza2018mobile}
    \end{quote}
From the examples given above, we see that Green \& Smith provide a recommendation for designing security \acp{API}. The recommendation can be applied to tools, API design, and general practice by developers who are integrating security with their applications. On the contrary, Mendoza~\etal{} offer a concrete recommendation. The recommendation is a policy expressed through WarDroid~\cite{mendoza2018mobile} and addresses API \emph{Construction}. Ancestries tell us a complex story between concrete recommendations that may be easier to validate, but often are \emph{standalone}, and broader recommendations that require a wide array of studies (over time) for validation. Furthermore, to devise a method for validating broader recommendations, one may need to refer back to the ancestry of these recommendations to understand the reason for their transformation over time.

The reason concrete recommendations may be easier to validate is due to their frequent association with a specific tool. This tool is presented as a solution that informs the recommendations the study has provided, therefore these recommendations can be empirically validated by validating the use of the tool. As a result, it is easier to see any direct effects and assess the impact of recommendations on the usability of security \acp{API}.

However, abstract recommendations are intentionally broader so that they can be applied to fields of software design and security. Studies that propose recommendations, based on the insights of older papers that present abstract recommendations, should dedicate their efforts to validating the recommendations through experimentation designed to measure the effectiveness on the usability of security \acp{API}. Such studies should also consider that ancestral recommendations, upon which they build, may not be strongly validated themselves.



The full chart of ancestor--descendant relationships, including empirical validation between papers is shown in the Appendix (Figure~\ref{fig:SLRAncestryMap}). This shows instances of empirical validation and the ancestor--descendant relationships between papers (and the recommendations they provide) across our corpus.

We focus on the 13 usable Security API designer papers and discuss how the majority of the recommendations they provide derive from Saltzer \& Schroeder~\cite{saltzer1975protection}, Bloch~\cite{bloch2001effective, bloch2006design}, and Gamma~\etal{}~\cite{gamma1993design}. This is an important finding because it shows that these works have stood the test of time and their recommendations have evolved through multiple works to become a strong influence on security API design recommendations today. Of further interest is that we find instances where recommendations from the 1975 paper of Saltzer \& Schroeder are directly referenced by security API design works of 2020.



\subsubsection{\mbox{Saltzer \& Schroeder: Once Upon A Time}}

In 1975, Saltzer \& Schroeder wrote the paper `The Protection of Information in Computer Systems' in which they presented 8 design principles to help guide the design of protection mechanisms and prevent security flaws~\cite{saltzer1975protection}. The design principles were \emph{adapted} by a revision of material originally published by Saltzer in 1974~\cite{saltzer1974protection}. Saltzer's earlier work from 1974 has been \emph{borrowed} by Schneider~\cite{schneider1999enforceable}, \emph{distilled} by Gong \& Ellison to influence Java platform security~\cite{Gong:2003:IJP:599797}, and \emph{compared} through a contemporary look at Saltzer \& Schroeder's design principles by Smith~\cite{smith2012contemporary}. Saltzer \& Schroeder's design principles have also been \emph{empirically} validated and \emph{borrowed}~\cite{siponen2000critical,denning1982cryptography}. These relationships can be seen through Figure~\ref{fig:SLRLanguageMiniMap} in the Appendix.

Saltzer \& Schroeder's work has been thoroughly influential through a range of relationships by works very different from each other, addressing fields from security policies~\cite{schneider1999enforceable} from security for Java applications~\cite{Gong:2003:IJP:599797} to cryptographic \acp{API}. This level of influence establishes Salzter \& Schroeder's work as a classic and a strong foundation for security API design recommendations.

Gutmann's release of the Cryptlib cryptographic API in 1995 acted as a gateway between the security engineering recommendations published by Saltzer \& Schroeder and 7 of our 13 security API designer seeds. Gutmann \emph{adapted} Saltzer \& Schroeder's design principles when designing Cryptlib~\cite{gutmann1995cryptlib}.

Gutmann's Cryptlib advertised a `high-level interface' and how abstraction can improve usability while maintaining a strong level of security.

\begin{quote}
  ``Cryptlib provides anyone with the ability to add strong security capabilities to an application in as little as half an hour, without needing to know any of the low-level details that make the encryption or authentication work.''~\cite{gutmann1995cryptlib}
\end{quote}
Bernstein~\etal{} build on the design of Cryptlib and presents NaCl, an even more abstracted cryptographic API that \emph{compared} NaCl to Cryptlib in detail~\cite{bernstein2012security}. NaCl, itself, has forks such as Libsodium which also has forks including Monocypher. We can see how strongly Gutmann's adaptation of Saltzer \& Schroeder's principles has influenced the design of cryptographic \acp{API} and applications today.

Later in 1999, Gutmann carried forward these adaptations when presenting his own set of recommendations to help improve the design of cryptographic security architecture~\cite{gutmann1999design}. Gutmann's recommendations were also an \emph{adaptation} of principles used to design NSA's Security Service API. Gutmann also \emph{compared} his recommendations to that used to design Microsoft's Crypto API. In 2002, Gutmann concludes his trilogy by presenting a set of recommendations in the form of lessons learned from implementing cryptographic software~\cite{gutmann2002cryptosoftware}. These recommendations are the oldest of the security API designer seeds. Compared to Saltzer \& Schroeder, Gutmann's recommendations~\cite{gutmann2002cryptosoftware} have not been as widely validated by or related to other works. The recommendations were \emph{distilled} in 2012 by Heninger~\etal{}~\cite{heninger2012mining} and later in 2016 by Green \& Smith~\cite{green2016developers}.

\begin{quote}
  ``Most crypto software is written with the assumption that the user knows what they’re doing, and will choose the most appropriate algorithm and mode of operation, carefully manage key generation and secure key storage, employ the crypto in a suitably safe manner, and do a great many other things that require fairly detailed crypto knowledge. However, since most implementers are everyday programmers .. the inevitable result is the creation of products with genuine naugahyde crypto.''~\cite{gutmann2002cryptosoftware}
\end{quote}


In 2016, Green \& Smith published a list of 10 recommendations to help developers create more usable and secure cryptographic \acp{API}~\cite{green2016developers}. The recommendations stemmed from the \emph{distillation} of Gutmann's work and the \emph{adaptation} of a series of API design recommendations defined by Bloch in 2006~\cite{bloch2006design}. Green \& Smith's work is the ancestor, in the ancestor--descendant relationship, to 4 security API designer seed papers. Their recommendations are also empirically validated by Patnaik~\etal{}~\cite{patnaik2019usability}, another security API designer seed paper.

Patnaik~\etal{}~\cite{patnaik2019usability} offered an \emph{empirical} validation in this chain through evaluating the 10 Green \& Smith~\cite{green2016developers} principles. Through an analysis of over 2400 Stack Overflow questions and responses from developers facing challenges using 7 cryptographic libraries, they found 16 usability issues which were mapped against the 10 principles of Green \& Smith~\cite{green2016developers}. They analyzed the extent to which the 10 principles encompassed the 16 usability issues and also identified additional issues that were not addressed by Green \& Smith's principles. Based on this, they derived additional recommendations: ~\emph{4 usability smells}, which are indicators that an interface may be difficult to use for its intended users.

In 2018 Mindermann~\etal{} presented recommendations for designing cryptographic libraries based on an experiment using Rust cryptographic \acp{API}~\cite{mindermann2018rust}.
They addressed insecure defaults, advertisement of authenticated encryption in low-level libraries, lack of warnings about deprecated/broken features, and the scarcity of documentation and example code from low-level libraries.
They \emph{compare} their set of recommendations against the 10 principles defined by Green \& Smith noting:
\begin{quote}
  ``Compared to Green \& Smith’s top ten principles, our recommendations are more specific but do not conflict with their suggestions.''~\cite{mindermann2018rust}
\end{quote}
Acar~\etal{}~\cite{acar2017comparing} \emph{adapted} Green \& Smith to evaluate participants' solutions from a controlled experiment in which 256 Python developers attempted tasks involving symmetric and asymmetric cryptography using one of five different cryptographic \acp{API}.

Oliveira~\etal{}~\cite{oliveira2018APIblindspots} \emph{distill} the work of Green \& Smith as well as Acar~\etal{} as part of an empirical study to understand the developer's perspective of API blindspots. Oliveira~\etal{} studied the developers ability to perceive blindspots through a series of code scenarios. Oliveira~\etal{} analyzed the developer's personal traits, such as; perception of correctness, familiarity with code, and level of experience.

Votipka~\etal{} aimed to understand what security errors developers tended to make and why. Votipka~\etal{} analyzed 94 submissions of code attempting security problems, and as a result labeled 182 unique security vulnerabilities~\cite{votipka2020securitymistakes}. Votipka's results served as an \emph{adaptation} of Green \& Smith's recommendations.

The recommendations published by Votipka~\etal{} in 2020 is the latest in an ancestral chain dating back to 1975 when Saltzer \& Schroeder presented a series of design principles aimed at protection mechanisms~\cite{saltzer1975protection}. Authors like Gutmann and Green \& Smith played a pivotal role when tailoring the design principles of Saltzer \& Schroeder towards the security API design recommendations of today~\cite{gutmann2002cryptosoftware, green2016developers}. However, it is interesting to identify a direct and contemporary relationship between Saltzer \& Schroeder and Votipka~\etal{}. Votipka \emph{borrowed} Saltzer \& Schroeder's principles to highlight design violations made by developers introducing too much complexity in their code. This shows two very different forms of evolution; on the one hand we see security engineering recommendations from 1975 strongly influential and transforming slowly over time to address challenges in niche fields such as the design of cryptographic \acp{API} and security API design recommendations, and on the other, we find Saltzer \& Schroeder's principles are still relevant today and flexible enough to address the challenges of designing security \acp{API} directly.


\subsubsection{\mbox{Bloch: Know Your Audience}}

In order to master a new language one must learn the grammar (how to correctly structure the language), the vocabulary (how to name things you want to talk about), and the common and effective ways in which to say things (usage). These practices are also applicable to programming languages. Many have addressed the first two practices thoroughly~\cite{arnold2000java, gosling2000java}. However, Bloch acknowledged that Java developers do not have a good understanding of the third practice---usage---and so, in 2001, Bloch dedicated Effective Java to address the practice of usage. The book offers advice on code structure, and the importance of others' understanding and code readability to improve ease of use when making future modifications~\cite{bloch2001effective}. Throughout the book, Bloch evaluates and compares his recommendations to Gamma~\etal{}'s design patterns~\cite{gamma1993design}. Bloch's ancestry can be fully traced through Figure~\ref{fig:SLRLanguageMiniMap} in the Appendix.

\begin{quote}
  ``A key feature of this book is that it contains code examples illustrating many design patterns and idioms. Where appropriate, they are cross-referenced to the standard reference work in this area [Gamma95]''~\cite{bloch2001effective}
\end{quote}
In 2001, Bloch takes a new direction and \emph{adapts} his recommendations from Effective Java to provide guidance for designing good \acp{API}. Initially Bloch provides this guidance in the form of a presentation at Google Inc. Following this, Bloch condenses the essence of the presentation into 39 recommendations. These recommendations were later adapted in 2016 by Green \& Smith to improve the usability of security \acp{API} through design.

Bloch's work is also \emph{adapted} by Acar~\etal{}, which \emph{adapts} the work of many, including Green \& Smith~\cite{green2016developers}, Henning~\etal{}~\cite{henning2007api}, and Nielsen~\cite{nielsen1994enhancing}, to compare the usability of Python based cryptographic \acp{API}~\cite{acar2017comparing}.

\begin{quote}
  ``We \emph{adapt} guidelines from these various sources to evaluate the \acp{API} we examine.''~\cite{acar2017comparing}
\end{quote}
We thus have an intricate chain of usable Security API designer recommendations---%
ones that both inform Green \& Smith~\cite{bloch2001effective,bloch2006design}, and those that Green \& Smith inform~\cite{acar2017comparing, patnaik2019usability, mindermann2018rust, oliveira2018APIblindspots, votipka2020securitymistakes}. In Bloch's ancestry, we do not find any explicit evidence that Bloch's recommendations have been empirically validated as they moved into Green \& Smith, though there is some traceability to Gamma~\etal{}'s design patterns\textemdash that are rooted in observations of developers' problem-solving practices. Bloch~\cite{bloch2001effective} discusses the many architectural advantages of Gamma~\etal{}'s design patterns, and more specifically Gamma~\etal{}'s factory pattern~\cite{gamma1993design}. This suggests that we need not only further empirical validation of Green \& Smith, but also upon the ancestry with which their work builds.

\subsubsection{\mbox{Georgiev: A Classic in the Making?}}

Georgiev~\etal{}'s work is in itself an empirical study, in that Georgiev~\etal{} provide evidence that the SSL certificate validation is broken in many security based applications and libraries. Georgiev~\etal{} found that any SSL connection from cloud clients based on the Amazon's E2C Java library are vulnerable to man-in-the-middle attacks. The reason for this issue are poorly designed \acp{API} of SSL implementations, in turn presenting developers with a confusing set of parameters and settings to decipher. Georgiev~\etal{} conclude their paper by presenting recommendations for both application developers and SSL library developers~\cite{georgiev2012mostdangerouscode}.

Georgiev~\etal{}'s recommendations are \emph{empirically} validated by O'Neill~\etal{}~\cite{oneill2018securesocketAPI}, another security API design seed, who also \emph{empirically} validate the works of Brubaker~\etal{}~\cite{brubaker2014using} and Fahl~\etal{}~\cite{fahl2012eve}. A connection is also seen between Georgiev~\etal{}, Brubaker~\etal{} and Fahl~\etal{}, as the latter two papers  \emph{compare} their work to that of Georgiev~\etal{}. This ancestry is shown through Figure~\ref{fig:SLRApplicationMiniMap} in the Appendix.

Building on Georgiev~\etal{} work, O'Neill~\etal{} present the Secure Socket API (SSA), a simplified TLS implementation using existing network applications~\cite{oneill2018securesocketAPI}. O'Neill~\etal{} build upon earlier work on TrustBase---an effort to improve security and flexibility available to administrators who select the certificate validation for their applications~\cite{oneill2017trustbase}. SSA presents the administrator with the choice of standard validation or TrustBase. By selecting TrustBase, administrators have finer-grained control over validation. O'Neill~\etal{} analyze the design of OpenSSL, providing recommendations to help improve the design. These recommendations generally apply when designing security \acp{API}.

Meng~\etal{} perform an empirical analysis of StackOverflow posts to understand challenges faced by developers when using secure coding practices in Java~\cite{meng2018securecodingpracticesjava}. They identify security vulnerabilities in the suggested code of answers provided through StackOverflow. The findings of the study suggests more consideration should be given to secure coding assistance and education, bridging the gap between security theory and coding practices. A comparison is made to Georgiev~\etal{}'s work and recommendations.

\begin{quote}
  ``\emph{Compared} with prior research, our study has two new contributions. First, our scope is broader. We report new challenges on secure coding practices, such as complex security configurations in Spring security, poor error messages, and multilingual programs. Second, our investigation on the online forum provides a new social and community perspective about secure coding. The unique insights cannot be discovered through analyzing code.''~\cite{meng2018securecodingpracticesjava}
\end{quote}

Similarly, Meng~\etal{} also \emph{compare} their work to Egele~\etal{}, who developed CryptoLint, a static program slicing tool designed to check applications from the Google Play marketplace. Egele~\etal{} find that 10,327 out of 11,748 applications that make use of cryptographic \acp{API} make at least one mistake. The criteria Egele~\etal{} based their analysis on was supported by two well-known security standards; ``security against chosen plaintext attacks (IND-CPA) and cracking resistance''~\cite{egele2013cryptomisuse}. Egele~\etal{} \emph{adapt} the work of Bellare~\etal{}~\cite{bellare2005introduction} and Desnos' Androguard~\cite{desnos2011androguard}.

\begin{quote}
  ``Our tool, called CryptoLint, is based upon the Androguard Android program analysis framework.''~\cite{egele2013cryptomisuse}
\end{quote}

\begin{quote}
  ``We adopt the notation used
by Bellare and Rogaway.''~\cite{egele2013cryptomisuse}
\end{quote}
Not only does the work of Bellare~\etal{} and Desnos provide a foundation for Egele~\etal{}'s analysis, but they directly influence the security criteria used by CryptoLint. Based on the analysis Egele~\etal{} present, a set of countermeasures against the vulnerabilities were found.

Between these 4 security API designer seeds, we can see that a good foundation is forming. In particular, Georgiev~\etal{}'s~\cite{georgiev2012mostdangerouscode} recommendations have been \emph{empirically} validated by O'Neill~\etal{}~\cite{oneill2018securesocketAPI}, and \textbf{\emph{compared}} by Meng~\etal{}~\cite{meng2018securecodingpracticesjava}, Brubaker~\etal{}~\cite{brubaker2014using}, and Fahl~\etal{}~\cite{fahl2012eve}. 
Does Georgiev~\etal{} have the makings of a classic? It may be so, as long as the recommendations continue to be empirically validated or related to by other ancestor--descendant relationships. It is essential to form a strong foundation upon which the community providing future security API recommendations and applications build.

\section{Threats to Validity}


We identify three main threats to validity.
First,
our search terms on Google Scholar and IEEExplore (see Section~\ref{sec:search-terms}) may have overlooked some papers relevant to our study. We mitigate this threat by manually reviewing:

\begin{itemize}
\item {IEEE Transactions on Software Engineering (TSE)}, \item {IEEE Symposium on Security \& Privacy (Oakland)}, \item {International Conference on Software Engineering (ICSE)},
\item {USENIX Security Symposium (USENIX)},  \item {International Symposium on Usable Privacy \& Security (SOUPS)}, and \item {ACM Conference on Computer and Communications Security (CCS)}.
\end{itemize}

We also used backward snowballing to trace the ancestry of the recommendations presented by our Security API designer seed papers, and we used forward snowballing on every paper along the ancestry to see if they were validated by other work. We have reviewed all the major conferences for papers that present recommendations relevant to our SLR and are confident that no relevant work has been overlooked.

Second,
the categorization was conducted inductively---%
meaning that our categories may not correlate with `common sense' understandings;
making comparison with other categorizations more difficult.
To mitigate this we calculated Cohen's $\kappa$~\cite{cohen1960coefficient}, demonstrating that our categorization was consistent between coders.
However, roughly a fifth of recommendations have two categories.
Cohen's $\kappa$ is not designed for data with multiple categories,
so when calculating inter-rater reliability we used only the first categorization.
This is unlikely to affect the overall analysis as the interrater reliability for just the first category is relatively high (0.74)---we would expect a second category to also be consistent.

Third,
we acknowledge that by looking at recommendations as the basis for the ancestry, we do not account for all ancestry of Security API designer papers.
We believe our analysis covers a significant proportion of the literature which is inherited by Security API designer papers and that, by noting the ways different papers have related to each other, we preserve how the knowledge developed.


\section{Discussion}

\subsection{The Classics}

What makes the works of Saltzer \& Schroeder, Bloch, Nielsen, and Gamma~\etal{} classics is that not only are they actionable, but also that they are flexible enough to transition into different branches of software engineering and computer security through adaptations~\cite{saltzer1975protection, bloch2001effective, bloch2006design, nielsen1994enhancing, gamma1993design}. This is no clearer than in Saltzer \& Schroeder's case~\cite{saltzer1975protection} where they define a set of recommendations to address the design challenges of protecting information, stored on computers, from unauthorized access. Gutmann facilitated the transition from security engineering to security API design by using Saltzer \& Schroeder's recommendations to design the Cryptlib cryptographic API~\cite{gutmann1995cryptlib}. Gutmann work is later related to by Green \& Smith, from which many other seed papers grew~\cite{green2016developers}. This evolution was possible primarily because Saltzer \& Schroeder's recommendations were actionable. The flexibility of Saltzer \& Schroeder's recommendations is tested again in 2020 directly by Votipka~\etal~\cite{votipka2020securitymistakes}, proving that not only have their recommendations stood the test of time but also that they are still relevant for addressing the challenges faced today by security API designers.

Gamma~\etal{} also play an influential role in the state of today's security API design recommendations. In 2001, Bloch transitions the design pattern's of Gamma~\etal{} to address usability challenges in Java programming~\cite{bloch2001effective}. In 2006, Bloch adapts his own work towards designing good APIs~\cite{bloch2006design}. Green \& Smith tailor Bloch's API design recommendations for security API designers~\cite{green2016developers}. The wide-spread influence seen through Gamma~\etal{}'s ancestry explains why it is a classic.

Saltzer \& Schroeder permeate several of our categories, but some also come from elsewhere
The \emph{Documentation} category likely has its origins in the work of Nielsen and UI~usability~\cite{nielsen1994enhancing}. Nielsen's recommendations are adapted by Acar~\etal{} for comparing the usability of Python based cryptographic APIs, where the importance of good documentation is highlighted.
Several other papers have highlighted the importance of usable and high quality documentation~\cite{bloch2006design,mindermann2018rust,nielsen1994enhancing,beaton2008usability,des2004eclipse,pane1996usability,zibran2008makes,robillard2009makes,patnaik2019usability,tondel2008security,bloch2001effective}.

Before going on to advise on how one can ease novice programmers into programming, Pane and Myers~\cite{pane1996usability} quote their guidance that:
\begin{quote}
    ``Even though it is better if the system can be used without documentation, it may be necessary to provide help and documentation. Any such information should be easy to search, focused on the user's task, list concrete steps to be carried out, and not be too large''~\cite{nielsen1994enhancing}
\end{quote}
Pane and Myer go on to inspire others and bring usability specifically to developers.
20 years later Green \& Smith describe their 10 principles for creating usable and secure crypto \acp{API}~\cite{green2016developers}.  Sure enough, one principle reads:
\begin{quote}
    ``Make \acp{API} easy to use, even without documentation''~\cite{green2016developers}
\end{quote}
Yet again, these arguably classic usability guidelines have been restated, rediscovered and then returned to.
However, there is no direct link in Green \& Smith to either Nielsen~\cite{nielsen1994enhancing} or Pane and Myers~\cite{pane1996usability}, demonstrating their pervasive role on `common sense'.
%


11 recommendations say specifically to use (and occasionally not to use~\cite{grill2012methods}) Gamma~\etal's \emph{Design Patterns} and 6 reference \emph{Factory Patterns} directly.
4 papers related to Gamma~\etal's work, and a further 2 validated the patterns empirically.
Perhaps the easy-to-recall names of many of the patterns have helped cement the work as a classic---but whilst we identified recommendations to use design patterns and to document their use, we did not see new versions of the \emph{Factory, Visitor, Observer} or \emph{Singleton patterns} being restated for Security API designer papers, or other more specialized fields.

This does not mean, however, that the original Gamma~\etal{}  \emph{Design Patterns} are not connected to the Security API designer field. The \emph{Design Patterns} have influenced the recommendations from the current literature (e.g., Bloch's), and have become an underlying standard upon which new recommendations are built.


The classics can be considered as a set of rules that are widely known to the current software engineering, computer security, and the usable security research communities. Future advances in security API design recommendations can refer to these standards, without hesitation, because the classics are tried and tested through developing challenges and time itself. 


\subsection{More Validation Please!}

Why is Saltzer \& Schroeder's work the only one to survive 1975? Why is a paper from 1994, authored by Nielsen, still influential today? Or why does a series of design patterns written by Gamma~\etal{} in 1993 part of an SLR written in 2020? The answer to all these questions is seen through empirical validation and our ancestor--descendant relationships. Without the ancestry chain stemming from Saltzer \& Schroeder, would Votipka even know their recommendations existed? It is unlikely, which is probably the case for many other design recommendation papers from that time. This is exactly why empirical validation is necessary. The purpose of empirical validation is to test the effectiveness of the recommendations suggested by a paper. Empirical validation helps set aside poor design recommendations and brings forward recommendations that prove to be effective. Empirical validation provides assurance to designers that the recommendations they are considering do in fact help design better software.

Whilst we found that many recommendations have not been validated or related to ($\frac{166}{883}$, 19\%), overall the software engineering and security communities seem to be making strides towards it.
Yet, this seems to be less so with papers that provide both general and security-focused recommendations for developing \acp{API}, at 39\% and 33\% of papers respectively (see Figure~\ref{tab:validation-counts}).
Our work shows that when we write research producing recommendations for developers, 22\% of all paper types are empirically validated.
More should be done to directly engage with the ancestral \emph{chains} deriving from older recommendations rather than validating a singular set of recommendations in order to ensure a depth from \emph{classics} to contemporary papers.
When creating new recommendations then, we should be looking at the history of where our knowledge comes from.

Therefore we argue that studies should not only focus on validating \emph{contemporary} papers, but also engage with the older and large body of knowledge concerning usability, and its implications for \acp{API} and security. In order to engage with older papers, one could run a standalone empirical validation study on the recommendations presented in these, for example. By applying recommendations in practice through experimentation and providing detailed analyses, one can help build more solid foundations as empirical validation can then be referenced by future studies. Upon this strong foundation, the recommendations can be transformed to create new recommendations specific to fields such as Security API designer guidance.

Many recommendations have arguably changed little from those made 25 or even 45 years ago---%
yet relatively few of the classics are referenced in the ancestral \emph{chains} we have analyzed.
Modern recommendations are clearly still being inspired by older works and to avoid restating ourselves, as a community we must take older, more established guidance and ensure---the foundational principles---are validated fully.

We see evidence of a classic in the making through the work of Georgiev~\etal{}~\cite{georgiev2012mostdangerouscode}. A set of recommendations that has been empirically validated and influenced the works of many others, Georgiev~\etal{} has the potential to influence many more in the field of security API design. To ensure this potential, more validation is needed, their recommendations need to be tested against varying conditions and challenges. 

\subsection{Meta-Recommendations}

Our categorization of the recommendations are neutral---%
we do not frame the categories as things one should or should not do---%
but rather describe what type of advice the recommendations offer.
After analyzing many different recommendations,
through a close reading and engagement, we offer meta-recommendations based on our extensive analysis of Security API designer papers. As we discussed previously, these are abstract and thus are not `actionable' by developers. However, these should guide broad thinking in both academic and practitioner material. These meta-recommendations are not exhaustive, but provide grounds for future thinking and development based on the findings from this SLR.

\begin{enumerate}
    \item \textbf{\itshape Do Quality Engineering~\cite{bloch2006design,beaton2008usability,nino2009introducing,bloch2001effective,assal2018security,MicrosoftSDL,BSIMM,OWASPSAMM,SeaCord2018SecurePractices}.} Software is not developed in isolation.  Have engineers and tools review code to spot rough edges and ensure best practice is followed.
    \item \textbf{\itshape Software Engineering Matters~\cite{bloch2006design,rivieres2004lines,Jacques2004APIGuidelines,sarkar2006api,nino2009introducing,gamma1993design,saltzer1975protection,saltzer1974protection}.} Follow best practices for software development and ensure code produced is of a high quality.  Give mechanisms for access control, and have a plan for how the code will be maintained.  Getting good, minimal, well abstracted, well structured code will pay dividends in the long run.
    \item \textbf{\itshape Embed Security at Every Stage~\cite{saltzer1975protection,CLASP2005ApplicationSecurity,lipner2004trustworthy,BSIMM}.} Design security in from the start by compartmentalizing components, and having sensible defaults.  Have a plan for dealing with bugs and defects.
    \item \textbf{\itshape Show, and Tell~\cite{bloch2006design,grill2012methods,patnaik2019usability,mindermann2018rust,tondel2008security}.} Documentation matters!  Document how the \acp{API} work. Document how programmers should use them. Provide exemplars. Standardize as much as possible.  Make sure the documentation is easy to find and read.
    \item \textbf{\itshape API Developers are not an Island~\cite{BSIMM,MicrosoftSDL,OWASPSAMM,CLASP2005ApplicationSecurity}.} An API might be for programmers to use, but they are often maintained and managed within organizations.  Executives need training to make good decisions, and organizations need a plan to develop their security knowledge and practices.
    API Developers will be influenced by outside forces (be they regulatory, risk-based, or third-party developers).
    \item \textbf{\itshape Write a Specification~\cite{sindre2005eliciting,meier2006web,haley2008security,mead2005security,SeaCord2018SecurePractices,assal2018security,MicrosoftSDL,BSIMM,bloch2006design}.} Start from requirements, and update those requirements as new threats are found.
    \item \textbf{\itshape Remember Programmers are Human~\cite{beaton2008usability,clarke2003using,stylos2007mapping,ko2004six,pane1996usability,grill2012methods,zibran2008makes,green2016developers,nielsen1994enhancing,nielsen1996usability,green1989cognitive,ko2005framework,molich1990improving,holcomb1989amalgamated,green1996usability}.} The first rule of code is you have to be able to read it.  Draw programmers' attention to the important bits; make it easy to spot mistakes, and to check when they have got it right.  Usability isn't just for users.
\end{enumerate}
 
\noindent
These 7 guiding principles summarize our 7 categories and bring together much of the advice in body of knowledge for developing secure \acp{API} as well as advice for more general software engineering.
They are not the sum total of all advice, but they cover what we distilled as a substantial amount with common points that multiple experts and papers have suggested.
We also note that some papers are referenced by many of the principles: \cite{bloch2006design,BSIMM,MicrosoftSDL,beaton2008usability,green2016developers} amongst others.
Perhaps then there should be an eighth principle:
\begin{enumerate}
    \setcounter{enumi}{7}
    \item \textbf{\itshape Know your Classics.} The struggles developers had when computers were first being programmed, and the strategies they came up with to deal with them, are still worth knowing about.  While more \emph{must} be done to validate these recommendations empirically, the refinement and restatement of them suggest they are still helpful.
\end{enumerate}


\section{Conclusion}

Our study is the first to systematically analyze 45 years of recommendations that inform Security API designer papers, crossing scientific communities working on security, \acp{API}, and software engineering. Our research questions systematize and learn where recommendations come from, whether they build on validated scientific work, and whether these bring a strong empirical focus to supporting developers with creating usable \acp{API}.

From an analysis of 65 papers guiding developers, including 13 specifically targeted at providing recommendations to developers on how to create usable and secure \acp{API}, and 883 recommendations found within the papers, we identified 7 broad categories of recommendations and 36 descriptor sub-categories.
These categories and descriptors provide a system for understanding the knowledge we have for guiding developers to produce better code, understand environments, and interface with organizations.
The community has made some strides towards validating recommendations, but more must be done within Security API designer literature to improve empirical validation.
As we identified, there are different types of ancestry according to their attention to \emph{abstract} and  \emph{concrete} recommendations.

Coverage is important alongside validation rates.
Through the ancestry analysis, we identified the well established ancestral chains between different areas of literature.
If the new Security API designer recommendations stem from a wider coverage of ancestral chains, it will be a stronger, more reliable set of Security API designer recommendations as more validation may have been carried out in the chain. This could result in more than one chain originating from historic sets of recommendations.
In addition, further developing work in the area ought to address the `classics' of the usability field in order to more appropriately attend to older principles and recommendations.
This is because, as we identify in our \emph{Meta-Recommendations}, many older and `classic' papers address similar contemporary recommendations.
Perhaps we don't need to reinvent the wheel so much as assess and renovate the parts to make them roadworthy for usable Security API designer recommendations today.

\balance
\bibliographystyle{IEEEtran}
\bibliography{bibliography}

\begin{thebibliography}{10}
\providecommand{\url}[1]{#1}
\csname url@samestyle\endcsname
\providecommand{\newblock}{\relax}
\providecommand{\bibinfo}[2]{#2}
\providecommand{\BIBentrySTDinterwordspacing}{\spaceskip=0pt\relax}
\providecommand{\BIBentryALTinterwordstretchfactor}{4}
\providecommand{\BIBentryALTinterwordspacing}{\spaceskip=\fontdimen2\font plus
\BIBentryALTinterwordstretchfactor\fontdimen3\font minus
  \fontdimen4\font\relax}
\providecommand{\BIBforeignlanguage}[2]{{%
\expandafter\ifx\csname l@#1\endcsname\relax
\typeout{** WARNING: IEEEtran.bst: No hyphenation pattern has been}%
\typeout{** loaded for the language `#1'. Using the pattern for}%
\typeout{** the default language instead.}%
\else
\language=\csname l@#1\endcsname
\fi
#2}}
\providecommand{\BIBdecl}{\relax}
\BIBdecl

\bibitem{robillard2009makes}
M.~P. Robillard, ``What makes {APIs} hard to learn? {A}nswers from
  developers,'' \emph{IEEE software}, vol.~26, no.~6, pp. 27--34, 2009.

\bibitem{mclellan1998building}
S.~G. McLellan, A.~W. Roesler, J.~T. Tempest, and C.~I. Spinuzzi, ``Building
  more usable {APIs},'' \emph{IEEE software}, vol.~15, no.~3, pp. 78--86, 1998.

\bibitem{kamp2014please}
P.-H. Kamp, ``Please put {OpenSSL} out of its misery.'' \emph{ACM Queue},
  vol.~12, no.~3, pp. 20--23, 2014.

\bibitem{nadi2016jumping}
S.~Nadi, S.~Kr{\"u}ger, M.~Mezini, and E.~Bodden, ``Jumping through hoops: Why
  do {Java} developers struggle with cryptography {APIs}?'' in
  \emph{Proceedings of the 38th International Conference on Software
  Engineering}.\hskip 1em plus 0.5em minus 0.4em\relax ACM, 2016, pp. 935--946.

\bibitem{gamma1993design}
E.~Gamma, R.~Helm, R.~Johnson, and J.~Vlissides, ``Design patterns: Abstraction
  and reuse of object-oriented design,'' in \emph{European Conference on
  Object-Oriented Programming}.\hskip 1em plus 0.5em minus 0.4em\relax
  Springer, 1993, pp. 406--431.

\bibitem{saltzer1975protection}
J.~H. Saltzer and M.~D. Schroeder, ``The protection of information in computer
  systems,'' \emph{Proceedings of the IEEE}, vol.~63, no.~9, pp. 1278--1308,
  1975.

\bibitem{brown2017finding}
F.~Brown, S.~Narayan, R.~S. Wahby, D.~Engler, R.~Jhala, and D.~Stefan,
  ``Finding and preventing bugs in {JavaScript} bindings,'' in \emph{2017 IEEE
  Symposium on Security and Privacy (SP)}.\hskip 1em plus 0.5em minus
  0.4em\relax IEEE, 2017, pp. 559--578.

\bibitem{green2016developers}
M.~Green and M.~Smith, ``Developers are not the enemy!: {The} need for usable
  security {APIs},'' \emph{IEEE Security \& Privacy}, vol.~14, no.~5, pp.
  40--46, 2016.

\bibitem{mendoza2018mobile}
A.~Mendoza and G.~Gu, ``Mobile application web {API} reconnaissance:
  Web-to-mobile inconsistencies \& vulnerabilities,'' in \emph{2018 IEEE
  Symposium on Security and Privacy (SP)}.\hskip 1em plus 0.5em minus
  0.4em\relax IEEE, 2018, pp. 756--769.

\bibitem{mindermann2018rust}
\BIBentryALTinterwordspacing
K.~Mindermann, P.~Keck, and S.~Wagner, ``How usable are rust cryptography
  {API}s?'' in \emph{2018 {IEEE} International Conference on Software Quality,
  Reliability and Security, {QRS} 2018, Lisbon, Portugal, July 16-20, 2018},
  2018, pp. 143--154. [Online]. Available:
  \url{https://doi.org/10.1109/QRS.2018.00028}
\BIBentrySTDinterwordspacing

\bibitem{patnaik2019usability}
N.~Patnaik, J.~Hallett, and A.~Rashid, ``Usability smells: An analysis of
  developers' struggle with crypto libraries,'' in \emph{Fifteenth Symposium on
  Usable Privacy and Security ({SOUPS} 2019)}, 2019.

\bibitem{bloch2006design}
J.~Bloch, ``How to design a good {API} and why it matters,'' in \emph{Companion
  to the 21st ACM SIGPLAN symposium on Object-oriented programming systems,
  languages, and applications}.\hskip 1em plus 0.5em minus 0.4em\relax ACM,
  2006, pp. 506--507.

\bibitem{bloch2001effective}
------, \emph{Effective Java}.\hskip 1em plus 0.5em minus 0.4em\relax Pearson
  Education, 2001.

\bibitem{stylos2007mapping}
J.~Stylos and B.~Myers, ``Mapping the space of {API} design decisions,'' in
  \emph{IEEE Symposium on Visual Languages and Human-Centric Computing (VL/HCC
  2007)}.\hskip 1em plus 0.5em minus 0.4em\relax IEEE, 2007, pp. 50--60.

\bibitem{nielsen1994enhancing}
J.~Nielsen, ``Enhancing the explanatory power of usability heuristics,'' in
  \emph{Proceedings of the SIGCHI conference on Human Factors in Computing
  Systems}.\hskip 1em plus 0.5em minus 0.4em\relax ACM, 1994, pp. 152--158.

\bibitem{beaton2008usability}
J.~Beaton, S.~Y. Jeong, Y.~Xie, J.~Stylos, and B.~A. Myers, ``Usability
  challenges for enterprise service-oriented architecture {API}s,'' in
  \emph{2008 IEEE Symposium on Visual Languages and Human-Centric
  Computing}.\hskip 1em plus 0.5em minus 0.4em\relax IEEE, 2008, pp. 193--196.

\bibitem{pane1996usability}
J.~F. Pane and B.~A. Myers, ``Usability issues in the design of novice
  programming systems,'' Carnegie-Mellon University, Tech. Rep., 1996.

\bibitem{zibran2008makes}
M.~Zibran, ``What makes {APIs} difficult to use,'' \emph{International Journal
  of Computer Science and Network Security}, vol.~8, no.~4, pp. 255--261, 2008.

\bibitem{tondel2008security}
I.~A. Tondel, M.~G. Jaatun, and P.~H. Meland, ``Security requirements for the
  rest of us: A survey,'' \emph{IEEE software}, vol.~25, no.~1, pp. 20--27,
  2008.

\bibitem{votipka2020securitymistakes}
D.~Votipka, K.~R. Fulton, J.~Parker, M.~Hou, M.~L. Mazurek, and M.~Hicks,
  ``Understanding security mistakes developers make: Qualitative analysis from
  build it, break it, fix it,'' in \emph{Proceedings of the 29 th USENIX
  Security Symposium (USENIX) Security}, vol.~20, 2020.

\bibitem{oneill2018securesocketAPI}
M.~O'Neill, S.~Heidbrink, J.~Whitehead, T.~Perdue, L.~Dickinson, T.~Collett,
  N.~Bonner, K.~Seamons, and D.~Zappala, ``The secure socket
  $\{$API$\}$:$\{$TLS$\}$ as an operating system service,'' in \emph{27th
  $\{$USENIX$\}$ Security Symposium ($\{$USENIX$\}$ Security 18)}, 2018, pp.
  799--816.

\bibitem{gutmann2002cryptosoftware}
P.~Gutmann, ``Lessons learned in implementing and deploying crypto software.''
  in \emph{Usenix Security Symposium}, 2002, pp. 315--325.

\bibitem{oliveira2018APIblindspots}
D.~S. Oliveira, T.~Lin, M.~S. Rahman, R.~Akefirad, D.~Ellis, E.~Perez,
  R.~Bobhate, L.~A. DeLong, J.~Cappos, and Y.~Brun, ``$\{$API$\}$ blindspots:
  Why experienced developers write vulnerable code,'' in \emph{Fourteenth
  Symposium on Usable Privacy and Security ($\{$SOUPS$\}$ 2018)}, 2018, pp.
  315--328.

\bibitem{acar2017comparing}
Y.~Acar, M.~Backes, S.~Fahl, S.~Garfinkel, D.~Kim, M.~L. Mazurek, and
  C.~Stransky, ``Comparing the usability of cryptographic {API}s,'' in
  \emph{2017 IEEE Symposium on Security and Privacy (SP)}.\hskip 1em plus 0.5em
  minus 0.4em\relax IEEE, 2017, pp. 154--171.

\bibitem{egele2013cryptomisuse}
M.~Egele, D.~Brumley, Y.~Fratantonio, and C.~Kruegel, ``An empirical study of
  cryptographic misuse in android applications,'' in \emph{Proceedings of the
  2013 ACM SIGSAC conference on Computer \& communications security}, 2013, pp.
  73--84.

\bibitem{georgiev2012mostdangerouscode}
M.~Georgiev, S.~Iyengar, S.~Jana, R.~Anubhai, D.~Boneh, and V.~Shmatikov, ``The
  most dangerous code in the world: validating ssl certificates in non-browser
  software,'' in \emph{Proceedings of the 2012 ACM conference on Computer and
  communications security}, 2012, pp. 38--49.

\bibitem{meng2018securecodingpracticesjava}
N.~Meng, S.~Nagy, D.~Yao, W.~Zhuang, and G.~A. Argoty, ``Secure coding
  practices in java: Challenges and vulnerabilities,'' in \emph{Proceedings of
  the 40th International Conference on Software Engineering}, 2018, pp.
  372--383.

\bibitem{henning2007api}
M.~Henning, ``{API} design matters,'' \emph{Queue}, vol.~5, no.~4, pp. 24--36,
  2007.

\bibitem{clarke2003using}
S.~Clarke and C.~Becker, ``Using the cognitive dimensions framework to evaluate
  the usability of a class library,'' in \emph{Proceedings of the First Joint
  Conference of EASE PPIG (PPIG 15)}, 2003.

\bibitem{robillard2011APIlearningobstacles}
M.~P. Robillard and R.~Deline, ``A field study of api learning obstacles,''
  \emph{Empirical Software Engineering}, vol.~16, no.~6, pp. 703--732, 2011.

\bibitem{rivieres2004lines}
J.~des Rivi{\`e}res, ``Eclipse {APIs}: Lines in the sand,'' \emph{EclipseCon
  Retrieved March}, vol.~18, p. 2004, 2004.

\bibitem{nino2009introducing}
J.~Ni{\~n}o, ``Introducing {API} design principles in {CS2},'' \emph{J. Comput.
  Small Coll}, vol.~24, no.~4, pp. 109--116, 2009.

\bibitem{grill2012methods}
T.~Grill, O.~Polacek, and M.~Tscheligi, ``Methods towards {API} usability: a
  structural analysis of usability problem categories,'' in \emph{International
  conference on human-centred software engineering}.\hskip 1em plus 0.5em minus
  0.4em\relax Springer, 2012, pp. 164--180.

\bibitem{ko2004six}
A.~J. Ko, B.~A. Myers, and H.~H. Aung, ``Six learning barriers in end-user
  programming systems,'' in \emph{2004 IEEE Symposium on Visual Languages-Human
  Centric Computing}.\hskip 1em plus 0.5em minus 0.4em\relax IEEE, 2004, pp.
  199--206.

\bibitem{myers2016improving}
B.~A. Myers and J.~Stylos, ``Improving {API} usability,'' \emph{Communications
  of the ACM}, vol.~59, no.~6, pp. 62--69, 2016.

\bibitem{cwalina2008framework}
K.~Cwalina and B.~Abrams, \emph{Framework design guidelines: conventions,
  idioms, and patterns for reusable .{NET} libraries}.\hskip 1em plus 0.5em
  minus 0.4em\relax Pearson Education, 2008.

\bibitem{green1989cognitive}
T.~R. Green, ``Cognitive dimensions of notations,'' \emph{People and computers
  V}, pp. 443--460, 1989.

\bibitem{green1996usability}
T.~R.~G. Green and M.~Petre, ``Usability analysis of visual programming
  environments: a `cognitive dimensions' framework,'' \emph{Journal of Visual
  Languages \& Computing}, vol.~7, no.~2, pp. 131--174, 1996.

\bibitem{kannampallil2006handling}
T.~G. Kannampallil and J.~M. Daughtry~III, ``Handling objects: a scenario based
  approach,'' in \emph{Proceedings of the 24th annual ACM international
  conference on Design of communication}.\hskip 1em plus 0.5em minus
  0.4em\relax ACM, 2006, pp. 92--98.

\bibitem{fowler2018refactoring}
M.~Fowler, \emph{Refactoring: improving the design of existing code}.\hskip 1em
  plus 0.5em minus 0.4em\relax Addison-Wesley Professional, 2018.

\bibitem{hoffman1990criteria}
D.~Hoffman, ``On criteria for module interfaces,'' \emph{IEEE transactions on
  Software Engineering}, vol.~16, no.~5, pp. 537--542, 1990.

\bibitem{molich1990improving}
R.~Molich and J.~Nielsen, ``Improving a human-computer dialogue,''
  \emph{Communications of the ACM}, vol.~33, no.~3, pp. 338--348, 1990.

\bibitem{holcomb1989amalgamated}
R.~Holcomb and A.~L. Tharp, ``An amalgamated model of software usability,'' in
  \emph{Proceedings of the Thirteenth Annual International Computer Software \&
  Applications Conference}.\hskip 1em plus 0.5em minus 0.4em\relax IEEE, 1989,
  pp. 559--566.

\bibitem{polson1990theory}
P.~G. Polson and C.~H. Lewis, ``Theory-based design for easily learned
  interfaces,'' \emph{Human-computer interaction}, vol.~5, no.~2, pp. 191--220,
  1990.

\bibitem{carroll1992getting}
J.~M. Carroll and M.~B. Rosson, ``Getting around the task-artifact cycle: how
  to make claims and design by scenario,'' \emph{ACM Transactions on
  Information Systems (TOIS)}, vol.~10, no.~2, pp. 181--212, 1992.

\bibitem{nielsen1996usability}
J.~Nielsen, ``Usability metrics: Tracking interface improvements,'' \emph{IEEE
  Software}, vol.~13, no.~6, pp. 1--2, 1996.

\bibitem{heninger2012mining}
N.~Heninger, Z.~Durumeric, E.~Wustrow, and J.~A. Halderman, ``Mining your ps
  and qs: Detection of widespread weak keys in network devices,'' in
  \emph{Presented as part of the 21st $\{$USENIX$\}$ Security Symposium
  ($\{$USENIX$\}$ Security 12)}, 2012, pp. 205--220.

\bibitem{Ko2004SoftwareErrorsFramework}
B.~A.~M. Andrew J.~Ko, ``A framework and methodology for studying the causes of
  software errors in programming systems,'' \emph{Visual Languages and
  Computing}, vol.~16, pp. 41--84, 2005.

\bibitem{Smith1982StarUI}
D.~C. Smith, C.~Irby, R.~Kimball, and B.~Verplank, ``{Designing the {Star} User
  Interface},'' \emph{BYTE}, pp. 242--282, April 1982.

\bibitem{OWASPDevGuide}
\BIBentryALTinterwordspacing
OWASP, ``{OWASP Developer Guide}.'' [Online]. Available:
  \url{https://www.owasp.org/index.php/Category: OWASP_Guide_Project}
\BIBentrySTDinterwordspacing

\bibitem{assal2018security}
H.~Assal and S.~Chiasson, ``Security in the software development lifecycle,''
  in \emph{Fourteenth Symposium on Usable Privacy and Security ({SOUPS} 2018)},
  2018, pp. 281--296.

\bibitem{MicrosoftSDL}
\BIBentryALTinterwordspacing
Microsoft, ``{Security Development Lifecycle}.'' [Online]. Available:
  \url{https://www.microsoft.com/en-us/sdl}
\BIBentrySTDinterwordspacing

\bibitem{BSIMM}
\BIBentryALTinterwordspacing
BSIMM. [Online]. Available: \url{https://www.bsimm.com}
\BIBentrySTDinterwordspacing

\bibitem{OWASPSAMM}
\BIBentryALTinterwordspacing
OWASP, ``{OWASP SAMM Project}.'' [Online]. Available:
  \url{www.owasp.org/index.php/OWASP_SAMM_Project}
\BIBentrySTDinterwordspacing

\bibitem{SeaCord2018SecurePractices}
\BIBentryALTinterwordspacing
R.~Seacord, ``{Top 10 secure coding practices}.'' [Online]. Available:
  \url{https://www.securecoding.cert.org/confluence/
  display/seccode/Top+10+Secure+Coding+Practices, 2011}
\BIBentrySTDinterwordspacing

\bibitem{fahl2012eve}
S.~Fahl, M.~Harbach, T.~Muders, L.~Baumg{\"a}rtner, B.~Freisleben, and
  M.~Smith, ``Why eve and mallory love android: An analysis of android ssl (in)
  security,'' in \emph{Proceedings of the 2012 ACM conference on Computer and
  communications security}, 2012, pp. 50--61.

\bibitem{ghafari2017security}
M.~Ghafari, P.~Gadient, and O.~Nierstrasz, ``Security smells in android,'' in
  \emph{2017 IEEE 17th international working conference on source code analysis
  and manipulation (SCAM)}.\hskip 1em plus 0.5em minus 0.4em\relax IEEE, 2017,
  pp. 121--130.

\bibitem{gutmann1999design}
P.~Gutmann, ``The design of a cryptographic security architecture.'' in
  \emph{USENIX Security Symposium}, 1999.

\bibitem{saltzer1974protection}
J.~H. Saltzer, ``Protection and the control of information sharing in
  {Multics},'' \emph{Communications of the ACM}, vol.~17, no.~7, pp. 388--402,
  1974.

\bibitem{mead2005security}
N.~R. Mead and T.~Stehney, \emph{Security quality requirements engineering
  ({SQUARE}) methodology}.\hskip 1em plus 0.5em minus 0.4em\relax ACM, 2005,
  vol.~30, no.~4.

\bibitem{haley2008security}
C.~Haley, R.~Laney, J.~Moffett, and B.~Nuseibeh, ``Security requirements
  engineering: A framework for representation and analysis,'' \emph{IEEE
  Transactions on Software Engineering}, vol.~34, no.~1, pp. 133--153, 2008.

\bibitem{bostrom2006extending}
G.~Bostr{\"o}m, J.~W{\"a}yrynen, M.~Bod{\'e}n, K.~Beznosov, and P.~Kruchten,
  ``Extending {XP} practices to support security requirements engineering,'' in
  \emph{Proceedings of the 2006 international workshop on Software engineering
  for secure systems}.\hskip 1em plus 0.5em minus 0.4em\relax ACM, 2006, pp.
  11--18.

\bibitem{CLASP2005ApplicationSecurity}
\emph{The {CLASP} Application Security Process}, Secure Software, Inc, 2005.

\bibitem{meier2006web}
J.~Meier, ``Web application security engineering,'' \emph{IEEE Security \&
  Privacy}, vol.~4, no.~4, pp. 16--24, 2006.

\bibitem{lipner2004trustworthy}
S.~Lipner, ``The trustworthy computing security development lifecycle,'' in
  \emph{20th Annual Computer Security Applications Conference}.\hskip 1em plus
  0.5em minus 0.4em\relax IEEE, 2004, pp. 2--13.

\bibitem{apvrille2005secure}
A.~Apvrille and M.~Pourzandi, ``Secure software development by example,''
  \emph{IEEE Security \& Privacy}, vol.~3, no.~4, pp. 10--17, 2005.

\bibitem{gorski2018developers}
P.~L. Gorski, L.~L. Iacono, D.~Wermke, C.~Stransky, S.~M{\"o}ller, Y.~Acar, and
  S.~Fahl, ``Developers deserve security warnings, too: On the effect of
  integrated security advice on cryptographic {API} misuse,'' in
  \emph{Fourteenth Symposium on Usable Privacy and Security ({SOUPS} 2018)},
  2018, pp. 265--281.

\bibitem{gutmann1995cryptlib}
P.~Gutmann, ``Peter gutmann. cryptlib security toolkit: user’s guide and
  manual,'' 1995.

\bibitem{wohlin2014snowball}
\BIBentryALTinterwordspacing
C.~Wohlin, ``Guidelines for snowballing in systematic literature studies and a
  replication in software engineering,'' in \emph{Proceedings of the 18th
  International Conference on Evaluation and Assessment in Software
  Engineering}, ser. EASE '14.\hskip 1em plus 0.5em minus 0.4em\relax New York,
  NY, USA: Association for Computing Machinery, 2014. [Online]. Available:
  \url{https://doi.org/10.1145/2601248.2601268}
\BIBentrySTDinterwordspacing

\bibitem{rivas2012coding}
C.~Rivas, ``Coding and analysing qualitative data,'' \emph{Researching society
  and culture}, vol.~3, pp. 367--392, 2012.

\bibitem{ando2014achieving}
H.~Ando, R.~Cousins, and C.~Young, ``Achieving saturation in thematic analysis:
  Development and refinement of a codebook,'' \emph{Comprehensive Psychology},
  vol.~3, pp. 03--CP, 2014.

\bibitem{cohen1960coefficient}
J.~Cohen, ``A coefficient of agreement for nominal scales,'' \emph{Educational
  and psychological measurement}, vol.~20, no.~1, pp. 37--46, 1960.

\bibitem{landis1977kappa}
\BIBentryALTinterwordspacing
J.~R. Landis and G.~G. Koch, ``An application of hierarchical kappa-type
  statistics in the assessment of majority agreement among multiple
  observers,'' \emph{Biometrics}, vol.~33, no.~2, pp. 363--374, 1977. [Online].
  Available: \url{http://www.jstor.org/stable/2529786}
\BIBentrySTDinterwordspacing

\bibitem{nielsen1990heuristic}
J.~Nielsen and R.~Molich, ``Heuristic evaluation of user interfaces,'' in
  \emph{Proceedings of the SIGCHI conference on Human factors in computing
  systems}.\hskip 1em plus 0.5em minus 0.4em\relax ACM, 1990, pp. 249--256.

\bibitem{smith2012contemporary}
R.~E. Smith, ``A contemporary look at {S}altzer and {S}chroeder's 1975 design
  principles,'' \emph{IEEE Security \& Privacy}, vol.~10, no.~6, pp. 20--25,
  2012.

\bibitem{schneider1999enforceable}
F.~B. Schneider, ``Enforceable security policies,'' Cornell University, Tech.
  Rep., 1999.

\bibitem{Gong:2003:IJP:599797}
L.~Gong and G.~Ellison, \emph{Inside {Java}\textsuperscript{TM} 2 Platform
  Security: Architecture, {API} Design, and Implementation}, 2nd~ed.\hskip 1em
  plus 0.5em minus 0.4em\relax Pearson Education, 2003.

\bibitem{siponen2000critical}
M.~T. Siponen, ``Critical analysis of different approaches to minimizing
  user-related faults in information systems security: implications for
  research and practice,'' \emph{Information Management \& Computer Security},
  vol.~8, no.~5, pp. 197--209, 2000.

\bibitem{denning1982cryptography}
D.~E.~R. Denning, \emph{Cryptography and data security}.\hskip 1em plus 0.5em
  minus 0.4em\relax Addison-Wesley Reading, 1982, vol. 112.

\bibitem{bernstein2012security}
D.~J. Bernstein, T.~Lange, and P.~Schwabe, ``The security impact of a new
  cryptographic library,'' in \emph{International Conference on Cryptology and
  Information Security in Latin America}.\hskip 1em plus 0.5em minus
  0.4em\relax Springer, 2012, pp. 159--176.

\bibitem{arnold2000java}
K.~Arnold, J.~Gosling, D.~Holmes, and D.~Holmes, \emph{The Java programming
  language}.\hskip 1em plus 0.5em minus 0.4em\relax Addison-wesley Reading,
  2000, vol.~2.

\bibitem{gosling2000java}
J.~Gosling, B.~Joy, G.~Steele, and G.~Bracha, \emph{The Java language
  specification}.\hskip 1em plus 0.5em minus 0.4em\relax Addison-Wesley
  Professional, 2000.

\bibitem{brubaker2014using}
C.~Brubaker, S.~Jana, B.~Ray, S.~Khurshid, and V.~Shmatikov, ``Using
  frankencerts for automated adversarial testing of certificate validation in
  ssl/tls implementations,'' in \emph{2014 IEEE Symposium on Security and
  Privacy}.\hskip 1em plus 0.5em minus 0.4em\relax IEEE, 2014, pp. 114--129.

\bibitem{oneill2017trustbase}
M.~O'Neill, S.~Heidbrink, S.~Ruoti, J.~Whitehead, D.~Bunker, L.~Dickinson,
  T.~Hendershot, J.~Reynolds, K.~Seamons, and D.~Zappala, ``Trustbase: an
  architecture to repair and strengthen certificate-based authentication,'' in
  \emph{26th $\{$USENIX$\}$ Security Symposium ($\{$USENIX$\}$ Security 17)},
  2017, pp. 609--624.

\bibitem{bellare2005introduction}
M.~Bellare and P.~Rogaway, ``Introduction to modern cryptography,'' \emph{Ucsd
  Cse}, vol. 207, p. 207, 2005.

\bibitem{desnos2011androguard}
A.~Desnos \emph{et~al.}, ``Androguard,'' 2011.

\bibitem{des2004eclipse}
J.~{des Rivi{\`e}res}, ``Eclipse {API}s: Lines in the sand,'' 2004, presented
  at EclipseCon.

\bibitem{Jacques2004APIGuidelines}
\BIBentryALTinterwordspacing
M.~Jacques, ``{API} usability: guidelines to improve your code ease of use,''
  2004, code Project. [Online]. Available:
  \url{http://www.codeproject.com/KB/usability/APIUsabilityArticle.aspx}
\BIBentrySTDinterwordspacing

\bibitem{sarkar2006api}
S.~Sarkar, G.~M. Rama, and A.~C. Kak, ``{API}-based and information-theoretic
  metrics for measuring the quality of software modularization,'' \emph{IEEE
  Transactions on Software Engineering}, vol.~33, no.~1, pp. 14--32, 2006.

\bibitem{sindre2005eliciting}
G.~Sindre and A.~L. Opdahl, ``Eliciting security requirements with misuse
  cases,'' \emph{Requirements engineering}, vol.~10, no.~1, pp. 34--44, 2005.

\bibitem{ko2005framework}
A.~J. Ko and B.~A. Myers, ``A framework and methodology for studying the causes
  of software errors in programming systems,'' \emph{Journal of Visual
  Languages \& Computing}, vol.~16, no. 1-2, pp. 41--84, 2005.

\end{thebibliography}
\vspace*{-2\baselineskip}
\begin{IEEEbiographynophoto}{Nikhil Patnaik} is a PhD student at University of Bristol.\end{IEEEbiographynophoto}
\begin{IEEEbiographynophoto}{Andrew C Dwyer} is an Addison Wheeler Research Fellow at the University of Durham.\end{IEEEbiographynophoto}
\begin{IEEEbiographynophoto}{Joseph Hallett} is a Senior Research Associate at University of Bristol.\end{IEEEbiographynophoto}
\begin{IEEEbiographynophoto}{Awais Rashid} is a Professor of Cybersecurity at University of Bristol.\end{IEEEbiographynophoto}

\clearpage
\onecolumn

\begin{figure}[H]\centering
\appendix
\includegraphics[width=0.5\linewidth]{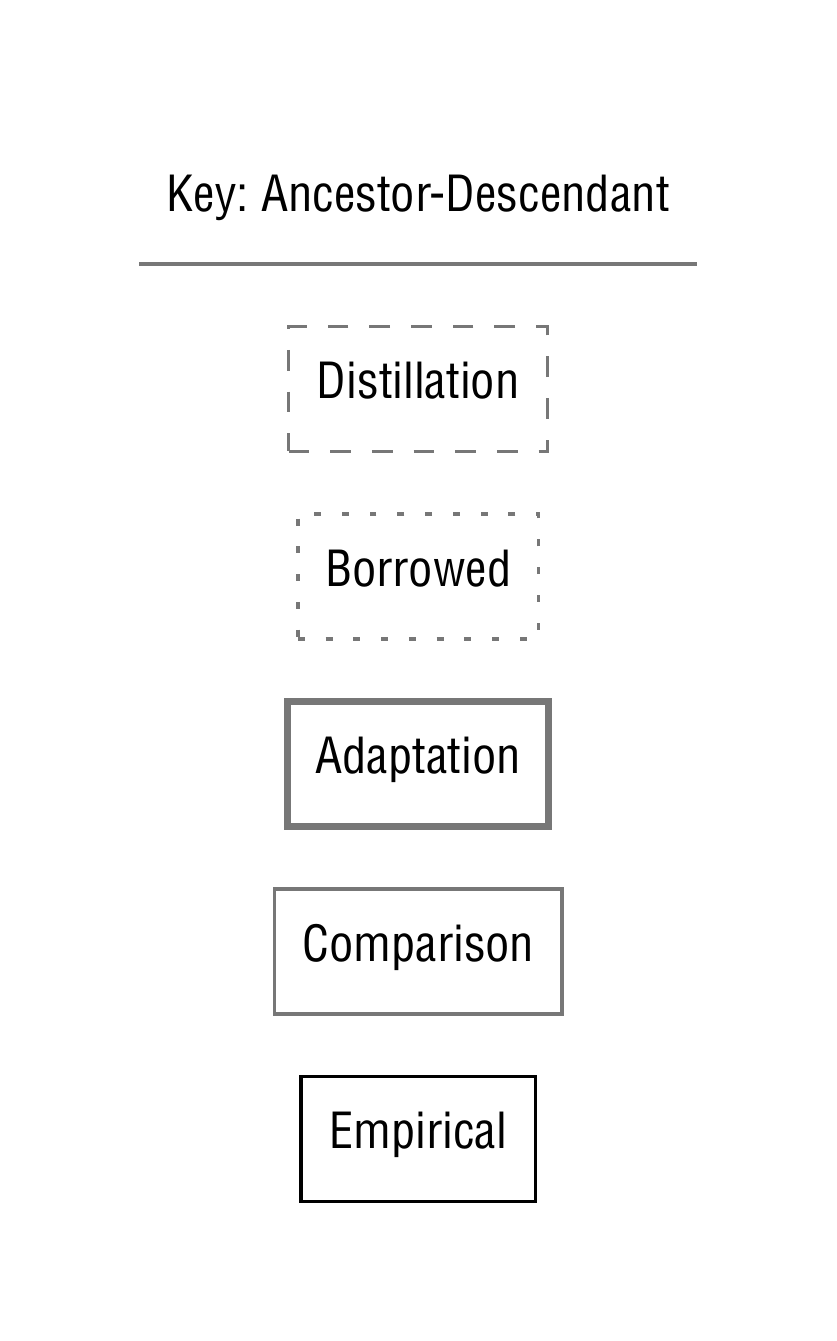}
\includegraphics[width=0.395\linewidth]{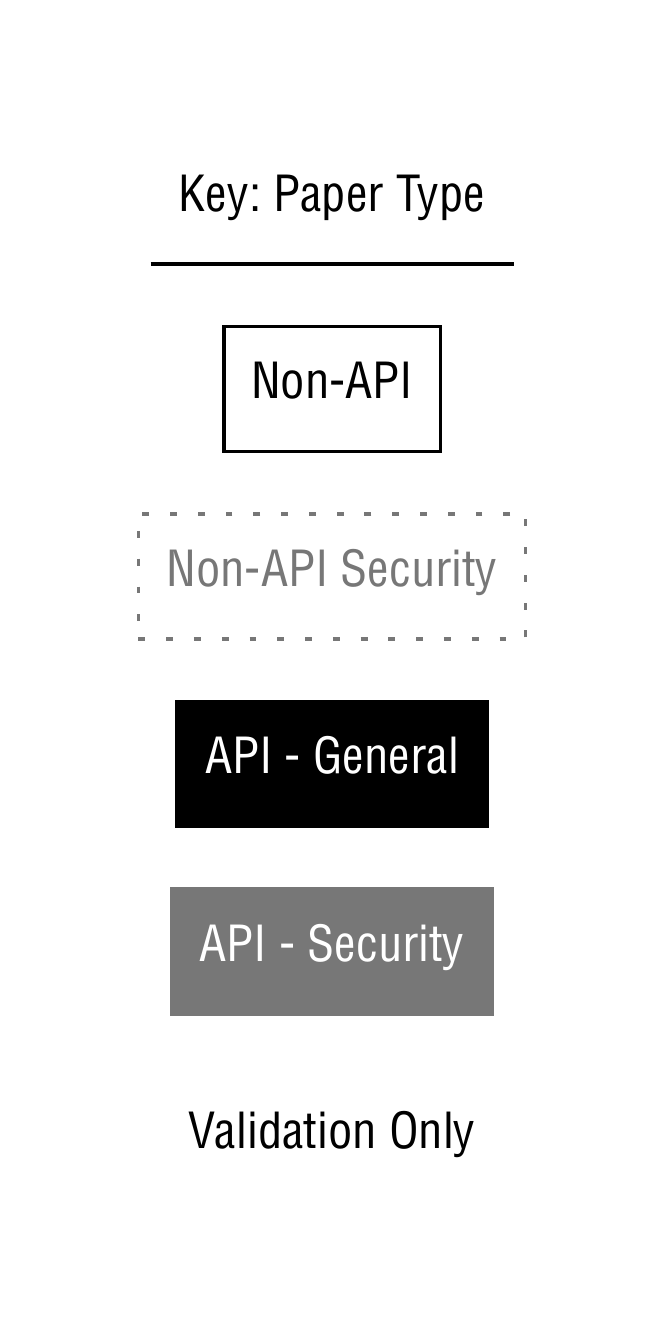}
\caption{Keys representing both Ancestor-Descendant relationships and types of papers. These keys help to guide through the following figures.
}
\label{fig:Legend}
\end{figure}

\newpage

\begin{figure}[H]\centering
\includegraphics[width=1.2\linewidth,angle=90]{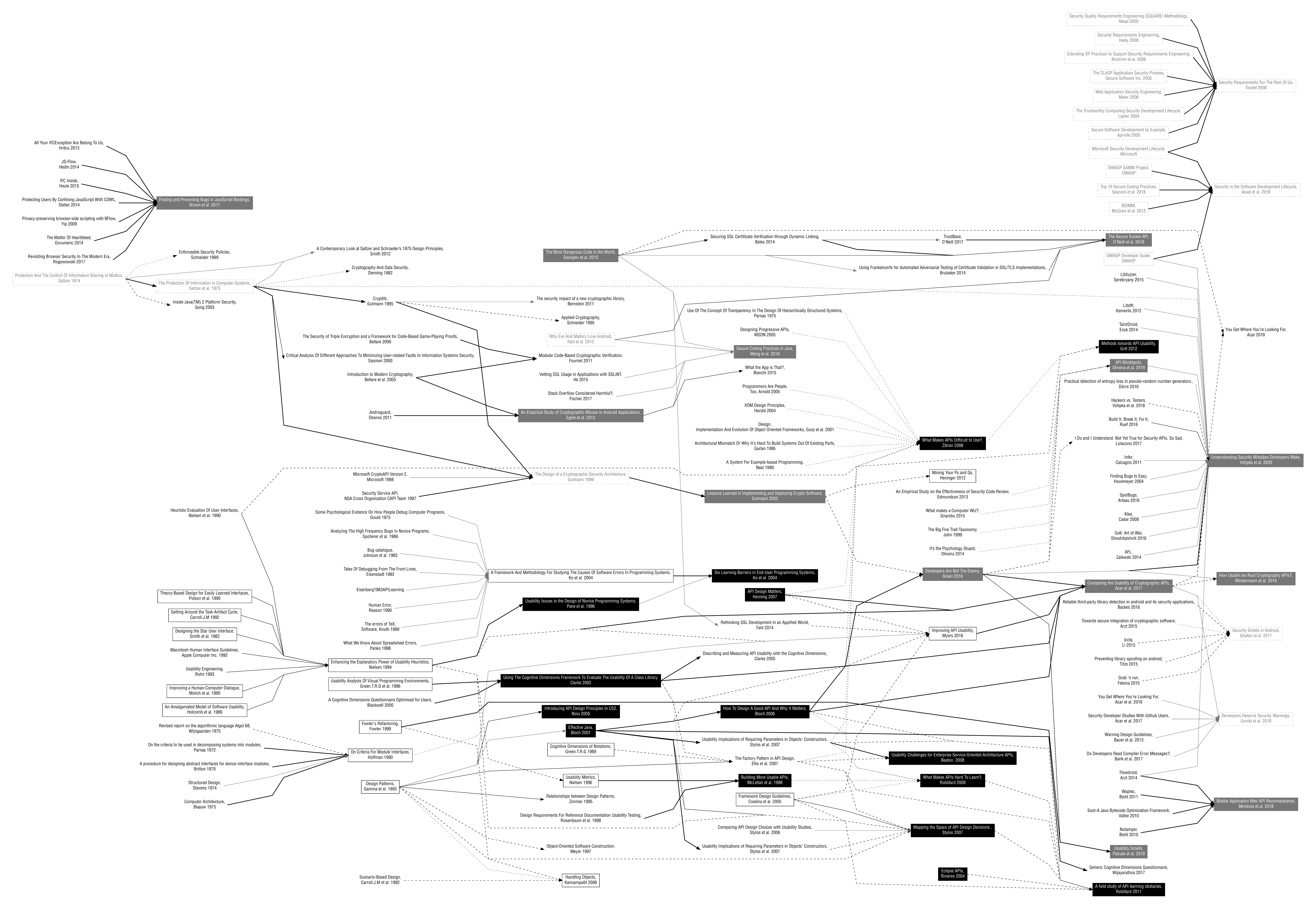}
\caption{Graph showing links between papers where one paper has built upon the work of another, and key. These were identified through the paper survey, using Security API designer papers with recommendations to identify and validate how knowledge has been translated.
}
\label{fig:SLRAncestryMap}
\end{figure}

\newpage

\begin{figure}
\centering
    \includegraphics[width=1.2\linewidth,angle=90]{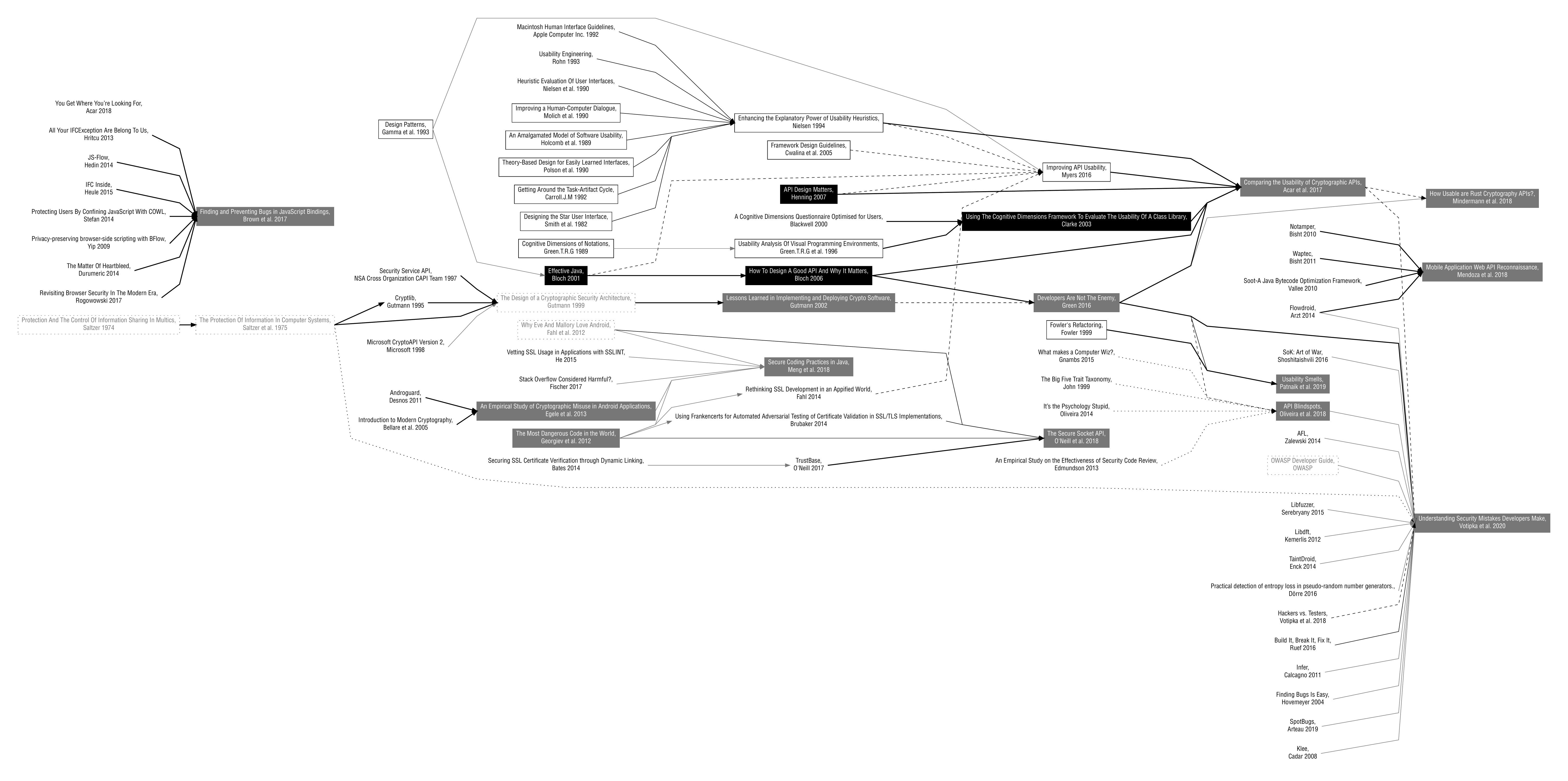}
    \caption{A filter version of Figure~\ref{fig:SLRAncestryMap}, with forward snowballing removed so that core ancestral chains are visible without additional validation seen through forward snowballing.}
\label{fig:SLRMiniMap}
\end{figure}

\newpage

\begin{figure}
\centering
    \includegraphics[width=1.2\linewidth,angle=90]{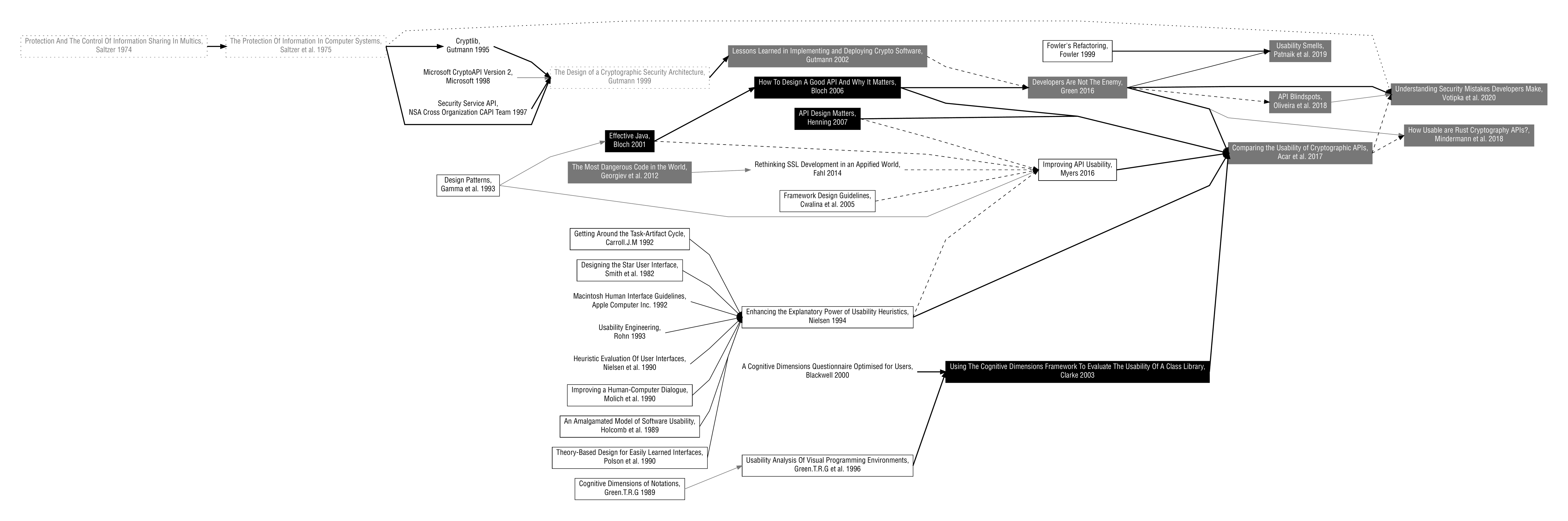}
    \caption{A map showing the ancestry surrounding Saltzer \& Schroeder~\cite{saltzer1975protection}, Bloch~\cite{bloch2001effective, bloch2006design} and Gamma~\etal{}~\cite{gamma1993design}.}
\label{fig:SLRLanguageMiniMap}
\end{figure}

\newpage

\begin{figure}
\centering
    \includegraphics[width=1.2\linewidth,angle=90]{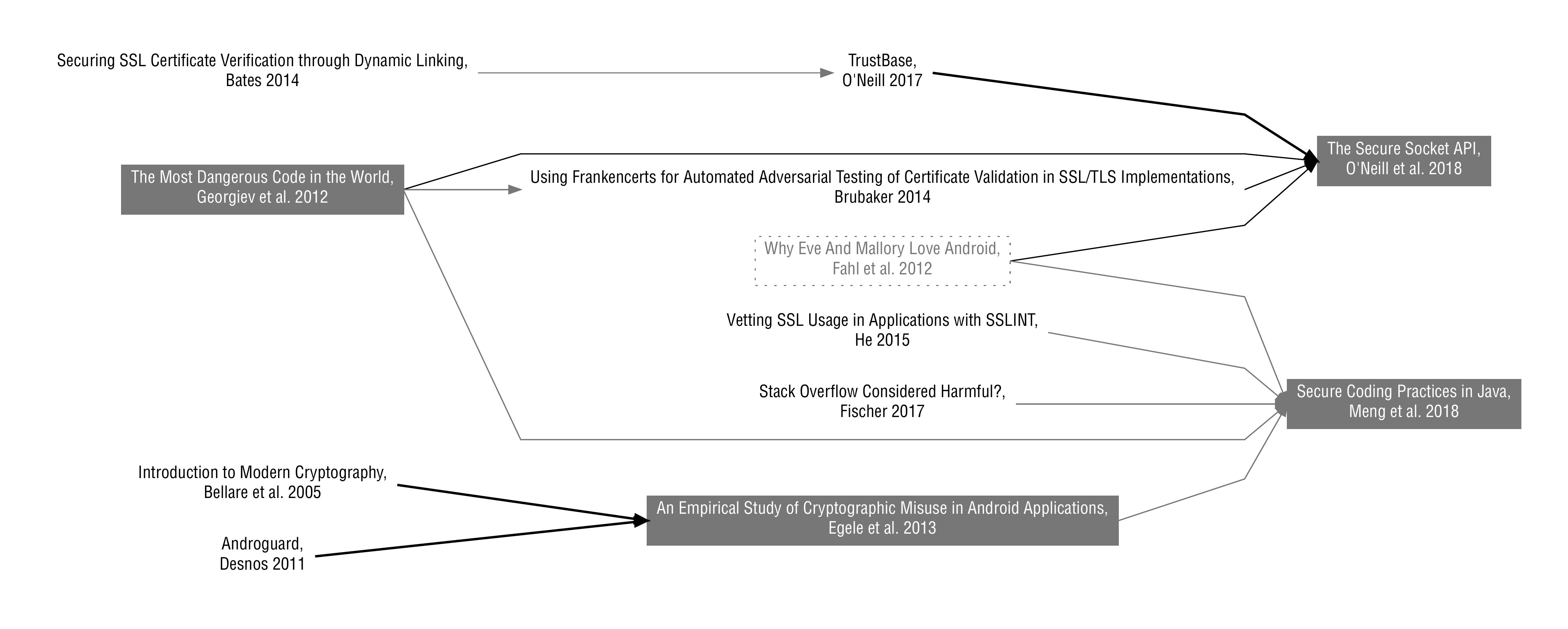}
    \caption{An ancestry originating from Georgiev~\etal{}~\cite{georgiev2012mostdangerouscode} is found isolated from Figure~\ref{fig:SLRLanguageMiniMap} but shares it's ancestry from O'Neill~\etal{}~\cite{oneill2018securesocketAPI}, Egele~\etal{}~\cite{egele2013cryptomisuse}, and Meng~\etal{}~\cite{meng2018securecodingpracticesjava}.}
\label{fig:SLRApplicationMiniMap}
\end{figure}

\newpage

\begin{figure}
\centering
    \includegraphics[width=1.2\linewidth,angle=90]{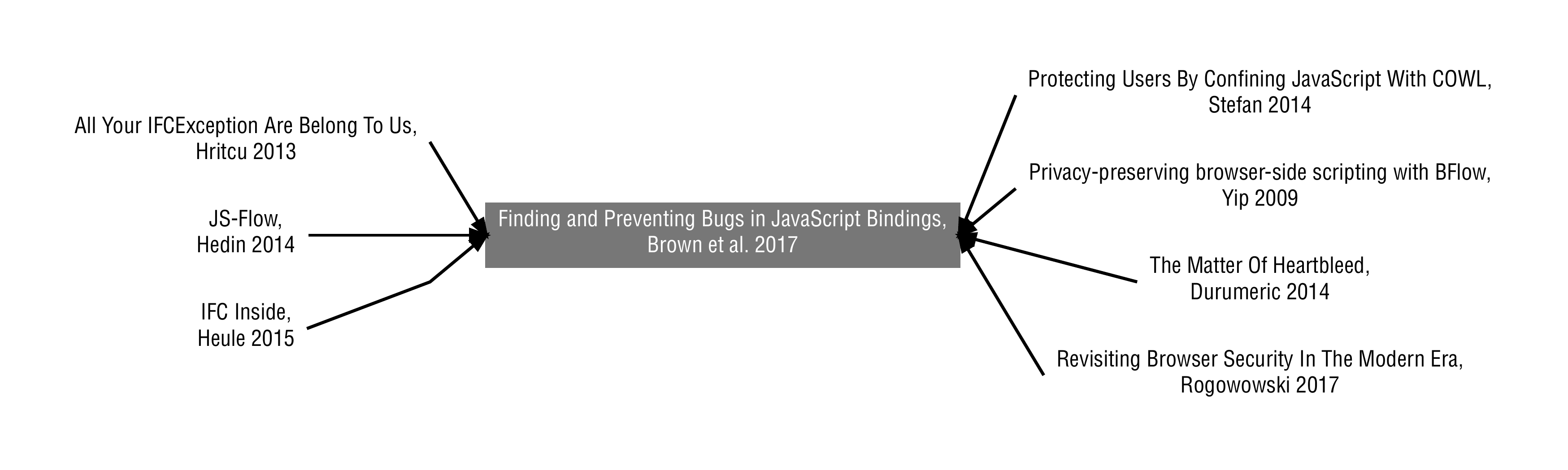}
    \caption{The isolated ancestry of Brown~\etal{}~\cite{brown2017finding}.}
\label{fig:SLRBrownMiniMap}
\end{figure}

\newpage

\begin{figure}
\centering
    \includegraphics[width=1.2\linewidth,angle=90]{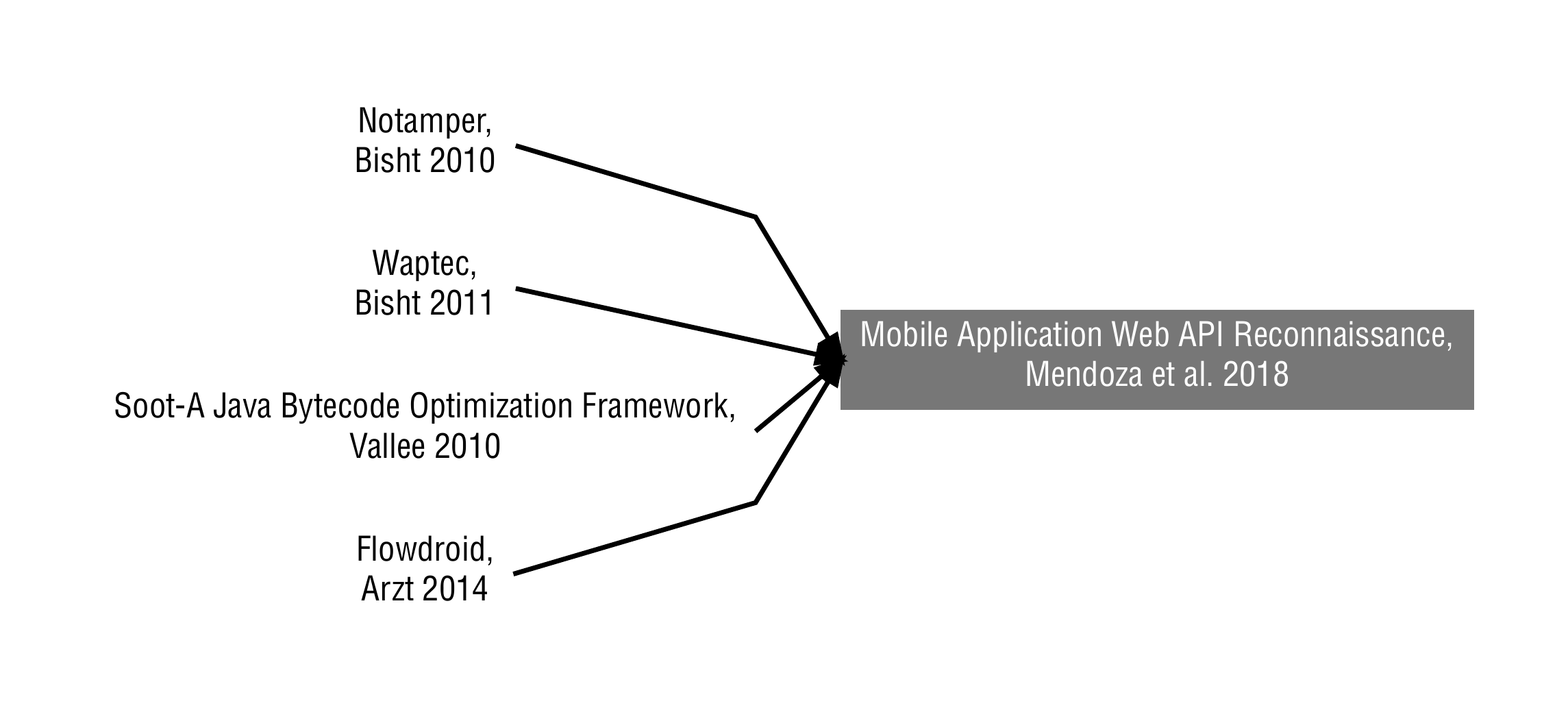}
    \caption{The isolated ancestry of Mendoza~\etal{}~\cite{mendoza2018mobile}.}
\label{fig:SLRMendozaMiniMap}
\end{figure}

\newpage

\begin{figure}
\centering
    \includegraphics[width=1.2\linewidth,angle=90]{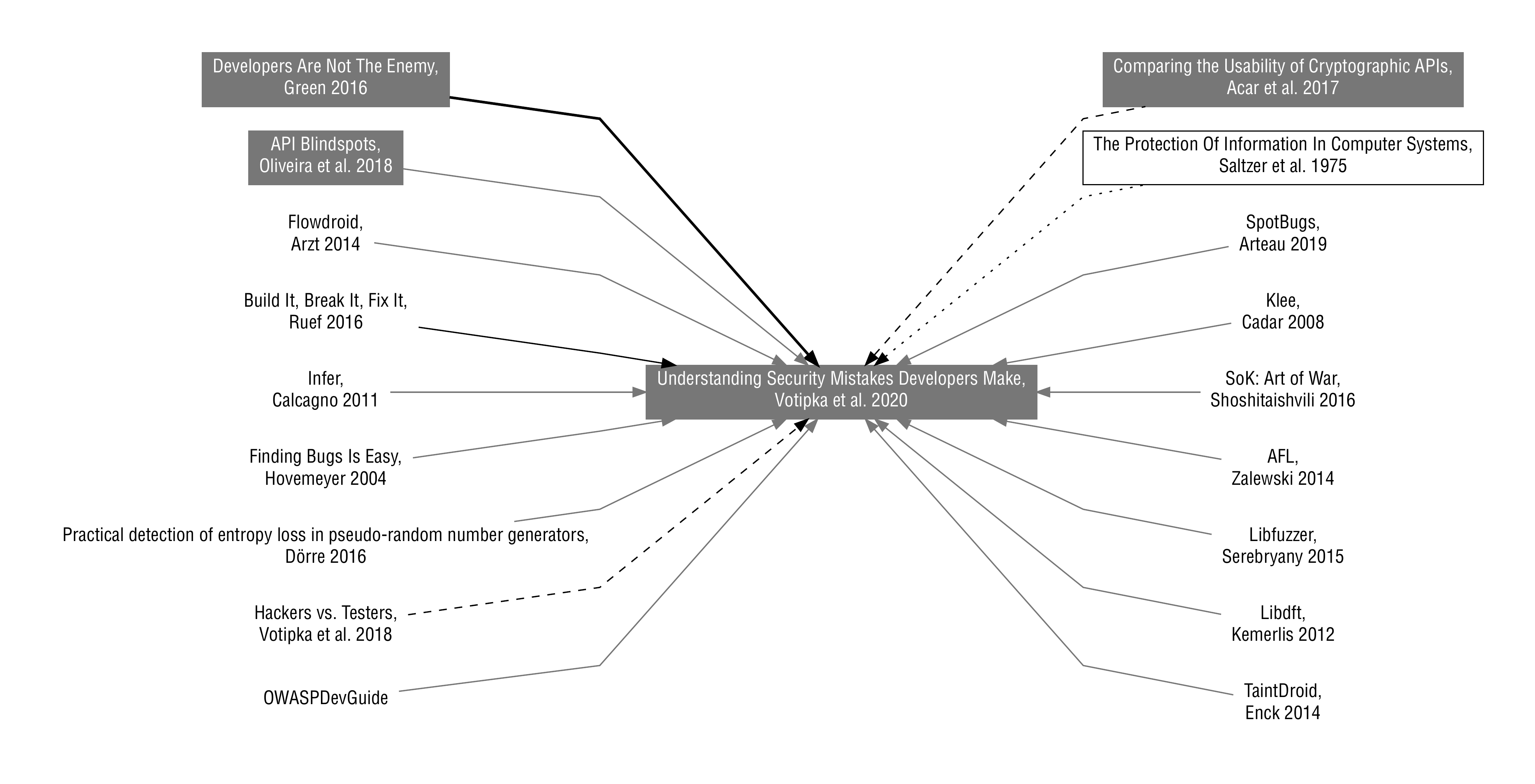}
    \caption{Votipka~\etal{}'s~\cite{votipka2020securitymistakes} ancestry is connected to Figure~\ref{fig:SLRLanguageMiniMap} through an adaptation of Green \& Smith's recommendations~\cite{green2016developers}. This figure presents the extended ancestry of Votipka~\etal{} through other relationships defined.}
\label{fig:SLRVotipkaMiniMap}
\end{figure}

\newpage

\begin{figure}
\centering
    \includegraphics[width=1.2\linewidth,angle=90]{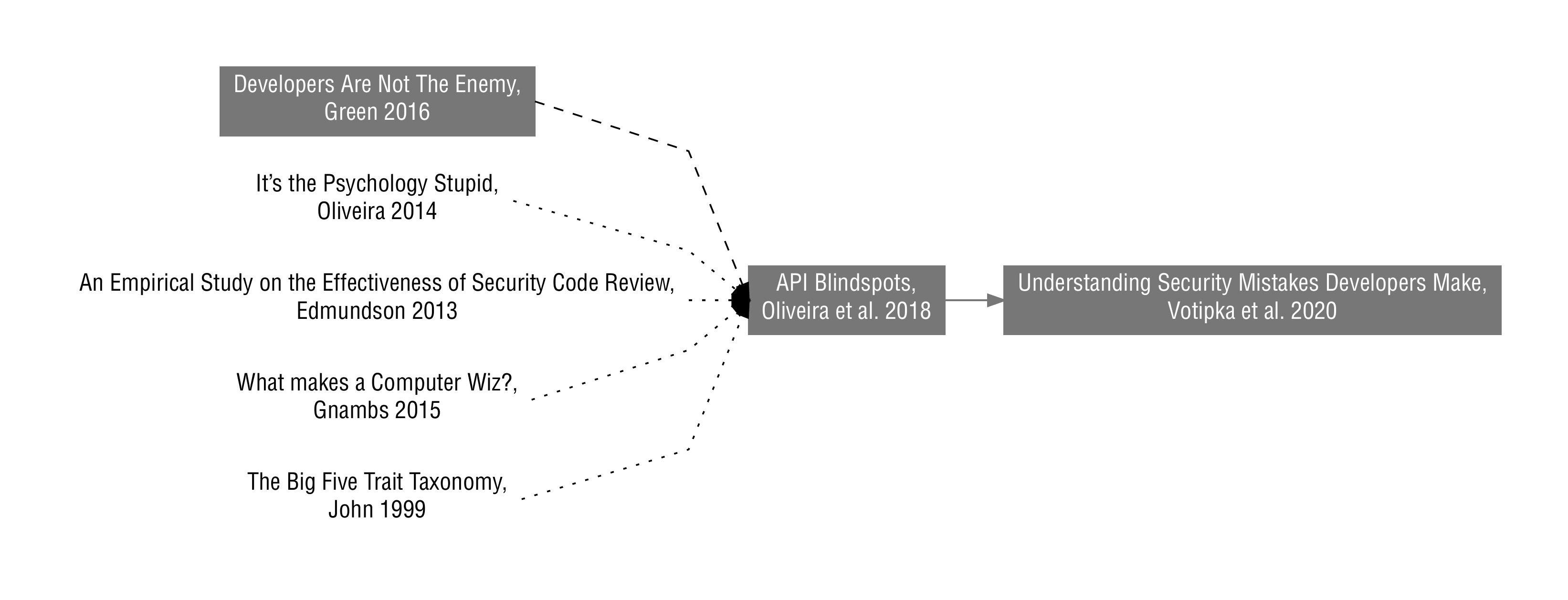}
    \caption{Oliveira~\etal{}'s~\cite{oliveira2018APIblindspots} ancestry is connected to Figure~\ref{fig:SLRLanguageMiniMap} through an adaptation of Green \& Smith's recommendations~\cite{green2016developers}. This figure presents the extended ancestry of Oliveira~\etal{} through other relationships defined.}
\label{fig:SLROliveiraMiniMap}
\end{figure}

\end{document}